\documentclass[npg]{copernicus}
\usepackage{multirow}
\begin{document}


\title{
Synchronization of coupled stick-slip oscillators}

\author[1]{N.~Sugiura}
\author[2]{T.~Hori}
\author[2]{Y.~Kawamura}

\affil[1]{Research Institute for Global Change, JAMSTEC, Yokosuka, Japan}
\affil[2]{Institute for Research on Earth Evolution, JAMSTEC, Yokohama, Japan}

\runningtitle{SYNCHRONIZED OSCILLATORS}

\runningauthor{SUGIURA ET AL.}

\correspondence{N.~Sugiura (nsugiura@jamstec.go.jp)}

\received{}
\pubdiscuss{} 
\revised{}
\accepted{}
\published{}


\firstpage{1}

\maketitle  

\begin{abstract}
A rationale is provided for the emergence of synchronization
in a system of coupled oscillators in stick-slip motion.
The single oscillator has a limit cycle in a region of the state space
for each parameter set beyond the supercritical Hopf bifurcation.
The two-oscillator system
that has similar weakly coupled oscillators
exhibits synchronization in a parameter range.
The synchronization has an anti-phase nature for an identical pair. However, it
 tends to be more in-phase for
a non-identical pair with a rather weak coupling.
A system of three identical
 oscillators (1, 2, and 3)
coupled in a line (with two springs $k_{12}=k_{23}$)
 exhibits synchronization with two of them (1 and 2 or 2 and 3) being nearly in-phase.
These collective behaviors are systematically estimated using
 the phase
reduction method.
\end{abstract}

\introduction  
Synchronization is ubiquitous in nature as there are numerous natural
networks of nonlinear dynamical systems \citep{pikovsky2003synchronization}.
Because faults that cause earthquakes or seismogenic processes can be
described as nonlinear dynamical systems, synchronization may occur in
fault behavior \citep{scholz2010large}.
The standard picture for the occurrence of interplate earthquakes is that a fault segment elastically driven by one plate, under the frictional resistance by another plate, exhibits a stick-slip motion that causes near-periodic spikes.
A group of such segments
can collectively cause recurring earthquakes
with some statistical regularity
\citep[e.g.,][]{scholz2002mechanics,kawamura2012statistical}. 
Although many factors about the interaction between fault segments are still unknown,
some evidence suggests that they can exhibit
synchronization \citep{rubeis2010synchronization}.
For example, \citet{chelidze2005} reported that
a stick-slip object
in a laboratory setting was entrained by a periodic force.
\citet{scholz2010large} statistically determined that the occurrence of earthquakes in some regions was clustered.
He reported that synchronous clusters of ruptures of
several faults were identified 
in the south Iceland seismic zone, the central Nevada seismic belt, and the eastern California shear zone.
 Meanwhile,
\citet{mitsui2004simple} successfully demonstrated that the numerically modeled coupled stick-slip oscillators
exhibited some degree of synchronization.
They used
a simple spring-slider system composed of
several mutually coupled stick-slip oscillators
to capture the nature of the
earthquake generation cycle along the Nankai trough,
which is located in a zone of high seismicity where multiple segments that constitute the fault zone have been reported to rupture almost simultaneously \citep{Ishibashi2004}.
It is worth noting that they found that a pair of coupled oscillators with slightly different parameter sets synchronized even for weak coupling
(Fig.\,6 of \citet{mitsui2004simple}),
although their emphasis was on cases with strong coupling between oscillators.

In spite of these observations, there has been little research that provides a specific description of the conditions for synchronization and how phases behave collectively. In this regard, we focus on the time evolution of the phases to elucidate the synchronization dynamics behind such collective behaviors and how phases are locked in the synchronization.

The occurrences of some earthquakes are nearly periodic \citep[e.g.,][]{matsuzawa2002characteristic,ishibashi2004seismotectonic,sykes2006repeat}; thus,
the generation process can be well modeled as a limit-cycle oscillation.
The timing of a limit-cycle oscillation can be
described by a single phase variable.
If the limit-cycles are somehow connected, they should interact with each other and exhibit some
collective behavior as a consequence
of the attraction or repulsion between
them in terms of the phase.
The phase reduction method \citep{kuramoto1984chemical}
enables us to quantify the rate at which
the progress of an oscillator phase
is affected by another oscillator,
thereby offering a powerful analytical tool
to approximate the limit-cycle dynamics
as a closed equation for
only a single phase variable.

We shall confine our attention to simple systems of
only a few oscillators that remain close to a common limit-cycle orbit,
rather than the complicated ones that
may produce chaotic motion \citep[e.g.,][]{huang1990evidence,HuangTurcotte1992,abe2012complex},
so that we can extract some regularity
from the collective behavior of the
oscillator system.
This setting, of assuming almost homogeneous system of limit cycle
oscillators, looks reasonable in the light of observations.
In fact, there are some seismic zones that consist of fault segments that have quite similar recurrence periods. 
The deviation of the earthquake generation 
periods between different segments along the Nankai trough 
is a few years, much smaller than the periods themselves,  
$\sim 1 \times 10^2 \mathrm{year}$ \citep{Ishibashi2004}.
Likewise, \citet{scholz2010large} points out that synchronization occurs within systems of evenly spaced, subparallel faults with very similar slip rates.

In this study, we quantitatively analyze how a single slider
oscillates under the rate- and state-dependent friction against a plate motion
using a bifurcation analysis and center-manifold reduction method.
Then, we identify when and how coupled sliders driven by a plate
synchronize as a collective substance using a phase reduction method.

\section{The spring-slider-dashpot system}
It is well established that a fault segment that
can cause earthquakes is well described
by a spring-slider system \citep[e.g.,][]{Perf2004}
subjected to a rate- and state-dependent friction
 \citep{dieterich1979modeling,ruina1983slip,scholz1998earthquakes};
this model exhibits a limit cycle oscillation.

Our research interest, therefore, is in spring-coupled sliders
(Fig.\,\ref{spring}) that are
driven by a common plate through spring and dashpot arrangements set for
each slider \citep[e.g.,][]{rice1993spatio,Cochard1994},
against the frictional resistance by another plate.
The equations of motion for the $i$-th slider are
\begin{eqnarray}
m_i \frac{d^2 x_i}{dt^2} &=&
k_i \left(V_p t - x_i - x_i^0 \right)
-\frac{G}{2c}\left( \frac{dx_i}{dt}-V_p \right)
-\tau_i
\nonumber\\
&&+\sum_j k_{ij} \left(x_j - x_i - x_{ij}^0 \right),\\
\frac{dx_i}{dt}&=&V_i,
\end{eqnarray}
where $x_i$ is the position of the slider,
$x_i^0$ and $x_{ij}^0$ are the lengths of springs at rest,
$k_i$ is the spring constant between the slider and plate,
$k_{ij}$ is the spring constant between a pair of sliders,
$V_i$ is the velocity,
$G$ is the rigidity,
$c$ is the shear wave velocity,
and $V_p$ is the constant velocity of the plate.
The frictional force $\tau_i$ has a rate- and state-dependent form that can be represented as \citep{ruina1983slip,dieterich1994direct}:
\begin{eqnarray}
\tau_i &=& \sigma_i \left( \mu^*_i + a_i
\log{\frac{V_i}{V^*}}+b_i\log{\frac{\theta_i}{\theta^*}} \right),
\end{eqnarray}
where
$a_i$ and $b_i$ are frictional parameters, $\sigma_i$ is the normal stress,
$V^*$ and $\theta^*$ are the arbitrary reference velocity and state, respectively, and
$\mu_i^*$ is a reference frictional coefficient.
The state variable $\theta_i$ obeys an aging law represented by
\citep{ruina1983slip,linker1992effects}:
\begin{eqnarray}
\frac{d\theta_i}{dt} &=& 1-\frac{V_i \theta_i}{L_i} \label{eq0_theta},
\end{eqnarray}
where $L_i$ is the characteristic length.
Under a quasi-static approximation where the
inertia $m_i d^2 x_i/dt^2 $
is sufficiently small \citep{gu1984slip,Perf2004,perfettini2005geodetic,Kano2010,Kano2013},
the governing equation for $V_i$
can be derived as
\begin{eqnarray}
\frac{d  V_i}{d t} &=& \frac{ k_{i}(V_p-V_i)-
 \frac{B_i}{\theta_i}\left( 1-\frac{V_i
		      \theta_i}{L_i}\right)}{\frac{A_i}{V_i}+g}\nonumber\\
&&+ \sum_j \frac{k_{ij}}{{\frac{A_i}{V_i}+g}} \left(V_j-V_i \right),
\label{eq1_V}
\end{eqnarray}
where $A_i=\sigma_i a_i$, $B_i=\sigma_i b_i$, and $g=G/(2c)$.
In accordance with the typical applications of the model to the seismogenic process, we
assume that the parameters are in the range of
\begin{eqnarray}
&&g>0,~ V_p>0, \\
(\forall{i})&& A_i>0,~ L_i>0,~ k_i>0,~ B_i- L_i k_i > 0,\\
(\forall{i \ne j})&&k_{ij} = k_{ji} \ge 0.
\end{eqnarray}
We also assume that all of the initial states are placed in the first quadrant:
\begin{equation}
(\forall{i})~ V_i(0)>0, ~\theta_i(0)>0.
\end{equation}

In Section\,\ref{sec:single}, we investigate the basic properties of a
single oscillator.
After introducing the phase reduction method in Section\,\ref{sec:prm},
we analyze the properties of synchronization, which
occurs in a two-oscillator system, in Section\,\ref{sec:pair}.
We mention some extensions to a three-oscillator system in Section\,\ref{sec:3-oscs}.

\section{The Dieterich-Ruina oscillator}\label{sec:single}
Here, we investigate the basic properties of a single
oscillator using the bifurcation and perturbation analyses.
\subsection{Governing equations}
Dropping the index $i$ in Eqs.\,(\ref{eq0_theta}) and (\ref{eq1_V})
for simplicity, we obtain the following equations describing a single oscillator:
\begin{eqnarray}
\frac{d  \theta}{d t} &=& 1-\frac{V \theta}{L},\label{eq_theta}\\
\frac{d  V}{d t} &=& \frac{ k(V_p-V)- \frac{B}{\theta}\left( 1-\frac{V\theta}{L}\right)}{\frac{A}{V}+g}.
\label{eq_V}
\end{eqnarray}
This is a two-dimensional dynamical system with six parameters
$(k, V_p, g, A, B, L)$.
Hereafter,
the dynamical system described by Eqs.\,(\ref{eq_theta}) and
(\ref{eq_V}) is called the Dieterich-Ruina oscillator,
and the state vector is denoted as
$\vec{X} = (\theta, V)^T$.
For the simplicity of analytical expressions, we use a new parameter set $(\mu, V_p, C, d, q, L)$
that is defined as
\begin{eqnarray}
 C &=& (A + g V_p)^{-1},~ d = C g V_p,~ q = \sqrt{CLk},\\
\mu &=& B-A-gV_p-Lk,
\end{eqnarray}
where $\mu$ serves as a bifurcation parameter.
Here, we investigate how the system behaves as $\mu$ changes.
This system has a unique equilibrium point at
$\vec{X}_0 = (L/V_p,V_p)^T$, which is given by the intersection of the nullclines:
\begin{eqnarray}
\mbox{I}: \quad V&=&\frac{L}{\theta} \quad \mbox{for}~ \frac{d \theta}{d t}=0,\\
\mbox{II}: \quad V&=&\left( \frac{B}{L}-k \right)^{-1} \left( \frac{B}{\theta}-k V_p
				      \right) \quad \mbox{for}~ \frac{d V}{d t}=0.
\end{eqnarray}
One of the important facts concerning the linear structure of the system around
an equilibrium point is that
the Jacobi matrix $\mathbf{J}$ has a characteristic sign pattern given by
\begin{equation}
\mathbf{J} = \left[\begin{array}{cc}
-\frac{V_p}{L} & -\frac{1}{V_p}\\
\frac{V_p^3 (q^2+1)}{L^2} & \frac{V_p}{L}
\end{array}\right]+\mu
\left[\begin{array}{cc}
0 & 0 \\
\frac{C V_p^3}{L^2} & \frac{C V_p}{L}
\end{array}\right]
=
\left[\begin{array}{cc}
- & -\\
+ & +
\end{array}\right],
\end{equation}
which represents a substrate-depletion system \citep{arcuri1986pattern}.
The two eigenvalues
of the Jacobi matrix at $\vec{X}_0$ are
\begin{equation}
\lambda_{1,2} = \frac{C V_p}{L}\mu \pm \frac{V_p}{2L}\sqrt{-4q^2+C^2
 \mu^2}.
\end{equation}

\subsection{Stable spiral: $\mu<0$}
The equilibrium point is a stable spiral when $-2q/C<\mu<0$
because the eigenvalues of $\mathbf{J}$ are
a complex conjugate pair,
\begin{equation}
\lambda_{1,2} = \frac{C V_p}{L}\mu \pm \mathrm{i} \frac{V_p}{2L}\sqrt{4q^2-C^2
 \mu^2},
\end{equation}
and have a common negative real part.

\subsection{Hopf bifurcation: $\mu=0$}
At the very instance when $\mu=0$,
the equilibrium point begins to lose its
stability.
The system encounters a Hopf bifurcation because
the Jacobi matrix has a pair of imaginary eigenvalues
\begin{equation}
\lambda_{1,2} = \pm  \mathrm{i}\frac{V_p q}{L}.
\end{equation}
The corresponding eigenvectors are
\begin{eqnarray}
\vec{U}=
\left[\begin{array}{c}
\frac{L(\mathrm{i} q-1)}{V_p(q^2+1)}\\
V_p
\end{array}\right]\label{def_U},
\end{eqnarray}
and its complex conjugate, $\overline{\vec{U}}$. $\vec{U}$ and $\overline{\vec{U}}$ span the plane containing linear solutions.
By introducing a complex amplitude, $W(t)$,
the neutral solution of the system
is expressed as
\begin{eqnarray}
\vec{X}(t)&=& \vec{X}_0 +\left\{ \vec{U} W(t) \exp{\left[\mathrm{i}\omega_0 t\right]}
			  +\mathrm{c.c.} \right\}, \label{neutral}\\
\omega_0 &=& \frac{V_p q}{L},\label{omega0}
\end{eqnarray}
where $\mathrm{c.c.}$ represents the complex conjugate.
The graph containing the solution indicates an elliptic orbital motion,
while the complex amplitude
is an arbitrary complex constant at this stage ($\mu=0$) 
if we neglect nonlinear terms.

\subsection{Weakly nonlinear: $\mu \gtrsim 0$}\label{sec_weakly_nonlinear}
When the bifurcation parameter $\mu$ becomes slightly larger than $0$,
the equilibrium point becomes an unstable spiral
because the eigenvalues of $\mathbf{J}$ are
a complex conjugate pair with a common positive real part.
Here, we develop an analytical expression for the asymptotic solutions in a weakly nonlinear regime, by expanding Eqs.\,(\ref{eq_theta}) and (\ref{eq_V}) as a Taylor series in terms of the deviation $\vec{u} \equiv \vec{X}-\vec{X}_0$ (See
appendix\,\ref{app_taylor}), and using $\vec{U}$ in Eq.\,(\ref{def_U}) and its dual, $\vec{U}^*$ (a left eigenvector), which is given by:
\begin{eqnarray}
\vec{U}^*&=&
\left(
-\mathrm{i} \frac{V_p (q^2+1)}{2 L q},~ V_p^{-1}\left(  \frac12 -
						\mathrm{i} \frac{1}{2q} \right)
\right).
\end{eqnarray}
With the expansion and eigenvectors, we can compute the coefficients for a small-amplitude equation
near the Hopf bifurcation following the center-manifold reduction method described in \citet{kuramoto1984chemical}.
Assuming the solutions are in the form of Eq.\,(\ref{neutral}),
the time evolution of the complex amplitude can be described by
the Stuart-Landau equation as
\begin{eqnarray}
\frac{d W}{d t}&=& \mu \alpha W - \beta \left|W\right|^2 W, \label{eq_weak_nonlin}\\
\alpha &=&
\vec{U}^* \mathbf{L}_1 \vec{U} =
\frac{C V_p}{2L},\label{alpha}\\
\beta
&=&
-3 \vec{U}^* \mathbf{N} \left( \vec{U}, \vec{U}, \overline{\vec{U}} \right) \nonumber \\
&&
+4 \vec{U}^*
\mathbf{M}
\left(
\vec{U}
,
\mathbf{M} (
\vec{U}, \mathbf{L}_0^{-1}\mathbf{M}(\vec{U},\overline{\vec{U}}
)
\right) \nonumber \\
&&
+2 \vec{U}^*
\mathbf{M} \left(
\overline{\vec{U}},
\left( \mathbf{L}_0 -2 \mathrm{i}\omega_0 \mathbf{I} \right)^{-1}
\mathbf{M}(\vec{U},\vec{U})
\right)
\nonumber \\
&=&
\frac{V_p}{2L}\left(d(1-d) + \mathrm{i}
\frac{q^2(1+2d)(1-d)+d^2}{3q}
\right).\label{beta}
\end{eqnarray}
This system encounters a supercritical bifurcation
to a stable limit cycle, because
the supercriticality condition $\mathrm{Re}\, \beta > 0$ is derived from
$0<d=g V_p/(A + g V_p)<1$.
Note that the type of the bifurcation
may have
some dependence on 
the laws of friction
and assumptions made on the equation of motion \citep{gu1984slip,Putelat201027}.
In the original vector form,
the limit-cycle solution of Eq.\,(\ref{eq_weak_nonlin}) is given by
\begin{eqnarray}
\vec{X}
&=& \vec{X}_0 +
\left\{
\vec{U} R_s
		       \exp{\left[
\mathrm{i} (\omega_0 + \tilde{\omega})t
\right]}
 + \mathrm{c.c.}
\right\}, \label{sol_SL} \\
R_s&=&
\sqrt{\frac{\mu \mathrm{Re}\,\alpha}{\mathrm{Re}\,\beta}}=
\sqrt{\frac{\mu C}{d (1-d)}},\label{rs}\\
\tilde{\omega} &=&
\mu \mathrm{Re}\,\alpha \left(
\frac{\mathrm{Im}\,\alpha}{\mathrm{Re}\,\alpha}
-
\frac{\mathrm{Im}\,\beta}{\mathrm{Re}\,\beta}
\right)\nonumber\\
&=&
-\mu \frac{CV_p}{L}\frac{q^2(1+2d)(1-d)+d^2}{6qd(1-d)},\label{omega1}
\end{eqnarray}
which graphically describes an elliptic orbital motion. 
The modulus, $R_s$, and frequency shift,
$\tilde{\omega}$, are scaled with $\mu^{1/2}$ and $\mu$, respectively.

We performed numerical integrations of Eqs.\,(\ref{eq_theta}) and
(\ref{eq_V})
to simulate the limit-cycle oscillation near the Hopf bifurcation point
for three cases with $\mu=10^{-5}$, $10^{-4}$,
and $10^{-3} \mathrm{Nm^{-2}}$.
The time integrations were performed with
the fourth-order Runge-Kutta scheme containing variable time step-sizes \citep{press1992numerical}.
The rest of the parameters were set according to a previous study by
\citet{Kano2010}
for an inter-plate earthquake occurred on September 25, 2003 in
Hokkaido, Japan:
$(V_p, g, A, B, L)=
(3.17 \times 10^{-9}\,\mathrm{ms^{-1}},
~ 5.00 \times 10^6\,\mathrm{Nm^{-3}s},
~ 1.50 \times 10^5\,\mathrm{Nm^{-2}},
~ 2.20 \times 10^5\,\mathrm{Nm^{-2}},
~ 1.00 \times 10^{-2}\,\mathrm{m})$;
these values also serve as the standard set of parameter values for this study.
In Fig.\,\ref{limitcycle0}, we show the results with the orbits of the limit cycle compared
to those derived using Eq.\,(\ref{sol_SL}).
The corresponding orbits are in good agreement when $\mu$ is small.

In the context of seismogenic processes, 
the analytical solution (Eq.\,(\ref{sol_SL})) in the weakly nonlinear regime 
may offer a simplified description of slow earthquakes
\citep[e.g.,][]{YoshidaKato2003,Heims2009}, which can be viewed as
sustaining aseismic oscillations in which the slip instability is
sufficiently weak \citep{kawamura2012statistical}.
In particular, the frequencies in Eqs.\,(\ref{omega0}) and (\ref{omega1})
can be used to evaluate the recurrence intervals of such earthquakes.

\subsection{Limit cycle: $\mu > 0$}
When we increase $\mu$, the system will enter a strongly nonlinear regime.
The equilibrium point becomes either
an unstable spiral (when $0<\mu<2q/C$) or
unstable node (when $\mu>2q/C$).
Then,
the Poincar\'{e}-Bendixson theorem \citep[e.g.,][]{strogatz2001nonlinear} ensures the existence of a limit
cycle within some region surrounding the equilibrium point,
because we now have an unstable equilibrium point
with a surrounding trapping region, $R$.
Appendix\,\ref{app_trap} describes how flows are trapped into the region.
Figure\,\ref{limitcycle} shows an example of a limit cycle orbit
derived by numerically integrating Eqs.\,(\ref{eq_theta}) and (\ref{eq_V}).
The orbit appears more polygonal than elliptical 
and extends over a wide range in the first quadrant.

\section{The phase reduction method}\label{sec:prm}
Here, we introduce the phase reduction method for general limit-cycle
oscillators, as well as
its specific representation for weakly nonlinear oscillators.
\subsection{Limit-cycle oscillators}
A system of coupled self-sustained oscillators can be
described by
\begin{eqnarray}
\frac{d \vec{X}_i}{d t} &=& \vec{F}(\vec{X}_i)+\delta
 \vec{f}_i(\vec{X}_i)+\sum_{j \neq i}\vec{g}_{ij}(\vec{X}_i, \vec{X}_j), \label{eq_Xi}
\end{eqnarray}
where we assume that the system $d\vec{X}/dt=\vec{F}(\vec{X})$ 
behaves by itself as a limit-cycle oscillator
and that the system described by Eq.\,(\ref{eq_Xi})
has an oscillatory behavior similar to it,
including the frequency and orbit.
Provided that the oscillators have similar properties and are weakly coupled, 
the phase reduction method \citep{kuramoto1984chemical}, shown below, is applicable to
the system.
Using the period, $T$, and the frequency, $\omega$, for
the limit cycle of the system $d\vec{X}/dt=\vec{F}(\vec{X})$,
we can define the phase, $\phi$, of a state that is
determined up to an integral multiple of $T$, which varies from $0$ to
$2\pi$. The time evolution of the phase obeys
\begin{eqnarray}
\frac{d \phi_i}{d t} &=& \omega + \delta \omega_i + \sum_{j \neq
 i} \Gamma_{ij}(\phi_i-\phi_j),
\end{eqnarray}
where $\phi_i$ is the phase of the oscillator $i$,
$\delta \omega_i$ is the frequency deviation of oscillator $i$ from
the original limit cycle frequency,
and $\Gamma_{ij}$ is the phase
coupling function (hereafter, the PCF)
between the oscillators $i$ and $j$, which is periodic with a period of $2\pi$.
These terms are defined as the averaged values of the deviation terms in
Eq.\,(\ref{eq_Xi})
over a period of the limit cycle
under the action of phase sensitivity, $\vec{Z}(\phi)$, (a row vector):
\begin{eqnarray}
\delta \omega_i &=& \frac{1}{2\pi}\int_0^{2\pi} \vec{Z}(\phi) \delta \vec{f}_i(\phi)\, d\phi,\label{eq_deltai}\\
\Gamma_{ij}(\psi)&=&
\frac{1}{2\pi}\int_0^{2\pi}\vec{Z}(\phi) \vec{g}_{ij}(\phi, \phi-\psi)\, d\phi.
\label{eq_gammai}
\end{eqnarray}
Here, $\vec{Z}(\phi)$ coincides with a left Floquet eigenvector, with eigenvalue 0, 
for the linearized equation around the limit cycle.
Refer to \citet{kuramoto1984chemical} for the details of the phase reduction method discussed here.

This procedure is applicable to the system containing Dieterich-Ruina oscillators
(Eqs.\,(\ref{eq0_theta}) and (\ref{eq1_V})),
provided that both 
the 
parameter differences 
and
coupling intensities 
of the oscillators
are small enough to be treated as a perturbation.
Substituting the specific functions in Eq.\,(\ref{eq1_V}) into Eq.\,(\ref{eq_gammai}),
we obtain the phase description of the system:
\begin{eqnarray}
\frac{d \phi_i}{d t} &=& \omega + \delta \omega_i + \sum_{j \neq
 i}k_{ij} \hat{\Gamma} \left( \phi_i-\phi_j \right),\label{phi_i}\\
\Gamma\left( \psi \right)&=&k_{ij}\hat{\Gamma}\left( \psi \right)\nonumber\\
&=&\frac{k_{ij}}{2\pi}\int_0^{2\pi}
\frac{V^*(\phi)}{A/V(\phi)+g}
\left[ V(\phi-\psi) - V(\phi) \right]\, d\phi,
\label{G_ij}
\end{eqnarray}
where $V^*$ is the phase sensitivity for $V$.
Note that $V$ and $V^*$ are defined along a stable orbit of a single
oscillator without coupling, which has a frequency $\omega$.

\subsection{Weakly nonlinear oscillators}
Suppose we have a system of weakly nonlinear oscillators that are
identical and mutually coupled.
Near the Hopf bifurcation point,
each oscillator can be described by
Eq.\,(\ref{eq_weak_nonlin}) and a coupling term,
which is supposed to be small:
\begin{eqnarray}
\frac{d  W_i}{d t} &=& \mu \alpha W_i - \beta |W_i|^2 W_i +
 \sum_{j \neq i}k_{ij}\gamma(W_j-W_i).
\end{eqnarray}
Normalizing the equations to
$t'=\left( \mu \mathrm{Re}\, \alpha \right) t$ and
$W'=\left( \mu \mathrm{Re}\, \alpha /\mathrm{Re}\,\beta
\right)^{-\frac12} W$, we get
\begin{eqnarray}
\frac{d  W'_i}{d t'} &=& (1+\mathrm{i}c_0) W'_i - \left( 1+ \mathrm{i} c_2 \right)|W'_i|^2 W'_i \nonumber\\
&&+\sum_{j \neq i} k'_{ij}(1+\mathrm{i} c_1)(W'_j-W'_i),\label{Wdash}\\
&&c_0= \frac{\mathrm{Im}\,\alpha}{\mathrm{Re}\,\alpha},\quad
c_1= \frac{\mathrm{Im}\,\gamma}{\mathrm{Re}\,\gamma},\nonumber \\
&&c_2= \frac{\mathrm{Im}\,\beta}{\mathrm{Re}\,\beta}, \quad
k'_{ij}=\frac{k_{ij}\mathrm{Re}\,\gamma}{\mu \mathrm{Re}\,\alpha}.\label{coe_Wdash}
\end{eqnarray}
By treating each oscillator as
a two-dimensional system with independent variables
$\left(\mathrm{Re}\,W'_i,\mathrm{Im}\,W'_i\right)^T$,
we can analytically derive the PCF
for this complex Ginzburg-Landau-type equation \citep{kuramoto1984chemical}:
\begin{eqnarray}
\Gamma_{ij}(\psi) &=&   -k'_{ij}\left[(1+c_1 c_2)\sin{\psi}+(c_2-c_1)(\cos{\psi}-1)\right]\label{eq:pcf_weak}.
\end{eqnarray}

For the case of the weakly nonlinear
Dieterich-Ruina oscillators,
(Eqs.\,(\ref{eq0_theta}) and (\ref{eq1_V})
 near the Hopf bifurcation point),
the coupling coefficient, $\gamma$, is defined in the same manner as $\alpha$ in Eq.\,(\ref{alpha}):
\begin{eqnarray}
 \gamma&=& \vec{U}^*
  \left[\begin{array}{cc}
 0 & 0\\
 0 & \frac{1}{A/V_p+g}
\end{array}\right]
\vec{U}=\left(\frac12 -
\mathrm{i} \frac1{2q} \right)
  \left( \frac{1}{A/V_p+g} \right).\label{KK}
\end{eqnarray}
Substituting Eqs.\,(\ref{alpha}),
(\ref{beta}), and (\ref{KK}) into Eq.\,(\ref{coe_Wdash}),
we get the coefficients in Eq.\,(\ref{Wdash}):
\begin{eqnarray}
&&c_0=0,\quad c_1= -\frac1q,\quad
 c_2=\frac{q^2
 (1+2d)(1-d)+d^2}{3qd(1-d)},\nonumber \\
&&k'_{ij}=\frac{k_{ij}L}{\mu} \geq 0.
\end{eqnarray}
Thus, the PCF
(Eq.\,(\ref{eq:pcf_weak}))
for the weakly nonlinear
Dieterich-Ruina oscillator is characterized by
\begin{eqnarray}
1 + c_1 c_2 &=& -\frac{q^2 (1-d)^2 + d^2}{3 q^2 d(1-d)}<0,\label{BF-ineq}\\
c_2 - c_1 &=& \frac{q^2 (1+2d)(1-d)+d(3-2d)}{3qd(1-d)}>0.
\end{eqnarray}
In particular, the inequality\,(\ref{BF-ineq}) indicates
that  the coupling has an anti-phase
nature ($d\Gamma/d\psi(0)>0,~d\Gamma/d\psi(\pi)<0)$).
Figure\,\ref{phase_coupling_weak} shows the PCF as a function of the phase
with the same parameters as in Fig.\,\ref{limitcycle0}.

\section{Two-oscillator system}\label{sec:pair}
Here, we explore when and how synchronization occurs in the system of two mutually coupled Dieterich-Ruina
oscillators.
We assume the two oscillators are identical except for a slight difference in the value of $B_i$.
To confirm the applicability of
the phase reduction method to the
stick-slip oscillator system, we examine the
properties of the synchronization in two different ways.
First, we observe the synchronization through
numerical integrations of a coupled oscillator system.
Second, we derive the PCF for the phase equations using the results from
the numerical integration of a single oscillator system and its
adjoint. Then, we determine some quantities from the plot.

\subsection{Numerical Integrations}\label{sec_direct}
We performed numerical integrations
of a discrete-time version of Eqs.\,(\ref{eq0_theta}) and
(\ref{eq1_V}),
for a pair of coupled oscillators:
a reference oscillator (oscillator 1)
and a second oscillator (oscillator 2).
Oscillator 1 had the following parameters:
$(k, V_p, g, A, B, L)=
(1.00 \times 10^5 \,\mathrm{Nm^{-3}},
~ 3.17 \times 10^{-9}\,\mathrm{ms^{-1}},
~ 5.00 \times 10^6\,\mathrm{Nm^{-3}s},
~ 1.50 \times 10^5\,\mathrm{Nm^{-2}},
~ 2.20 \times 10^5\,\mathrm{Nm^{-2}},
~ 1.00 \times 10^{-2}\,\mathrm{m})$,
and the natural frequency was $1.0687876 \times 10^{-9} \mathrm{s^{-1}}$. 
Two identical oscillators are coupled for case 0.
For cases 1, 2, and 3, we used oscillator 2 that has
the same set of parameters as oscillator 1 except for
$B=2.2025 \times 10^5$, $2.225 \times 10^5$, and $2.25 \times 10^5\mathrm{Nm^{-2}}$, respectively.
The natural frequencies of oscillator 2 in cases 1, 2, and 3 were
$1.0652082 \times 10^{-9} \mathrm{s^{-1}}$,
$1.0340304 \times 10^{-9} \mathrm{s^{-1}}$,
and
$1.00143857 \times 10^{-9} \mathrm{s^{-1}}$, respectively.
We used a common coupling strength of $K=k_{12}=k_{21}=3\times 10^3\,\mathrm{Nm^{-3}}$
for all cases, 
based on the one used in \citet{Kano2010}, 
which was derived through the inversion of strain rate from the GPS observation.
We also checked that the values of $B$ and $K$ were within the range of application of the phase reduction method (See appendix \ref{app_app}).
The time integrations are performed using the same method described in Section
\ref{sec_weakly_nonlinear}.
Figure\,\ref{V_case0} shows the results of case 0,
in which the oscillators synchronize at the phase difference
$\psi=-3.14$ (anti-phase).
Figure\,\ref{V_case20_00006} shows the results of case 1,
in which the oscillators synchronize at the phase difference
$\psi=-1.18$ (out-of-phase).
Figure\,\ref{V_case2_00006} shows the results of case 2,
in which
the oscillators synchronize at the phase difference $\psi=-7.07\times
10^{-3}$ (almost in-phase).
Figure\,\ref{V_case2_00003} shows the results of case 3,
in which they exhibit no synchronization.
These phase differences and synchronized oscillator frequencies are
listed in Table\,\ref{sync_prop}.

\subsection{Application of the PCF}\label{sec_PCF}
In this setting, the evolution of the phases can be described as
 \begin{eqnarray}
 \frac{d \phi_1}{d t} &=& \omega + \delta \omega_1 +
 K \hat{\Gamma} \left( \phi_1-\phi_2  \right), \label{eq_2elem1}\\
 \frac{d \phi_2}{d t} &=& \omega + \delta \omega_2 +
 K \hat{\Gamma} \left( \phi_2-\phi_1  \right), \label{eq_2elem2}
 \end{eqnarray}
where the PCF is defined in
Eq.\,(\ref{G_ij}), and the difference
between the natural frequencies is estimated to be
\begin{eqnarray}
\Delta \omega &\equiv & \delta \omega_1 - \delta \omega_2\\
&=&
 \frac{B_1-B_2}{2\pi}\int_0^{2\pi}
\frac{-V^*(\phi)/\theta(\phi)}{A/V(\phi)+g}\left( 1-\frac{V(\phi)
				    \theta(\phi)}{L} \right)\,d\phi \label{eq_domega}.
\end{eqnarray}

Taking the difference between Eqs.\,(\ref{eq_2elem1}) and (\ref{eq_2elem2}),
we
obtain the time evolution
of the phase difference, $\psi=\phi_1-\phi_2$:
\begin{eqnarray}
\frac{d \psi}{d t}    &=& 2K \left[ \frac{\Delta \omega}{2K} + \hat{\Gamma}_a(\psi) \right],\label{anti-PCF}\\
\hat{\Gamma}_a(\psi)&\equiv&\frac12\left( \hat{\Gamma}(\psi) - \hat{\Gamma}(-\psi)\right).\label{def_anti}
\end{eqnarray}
Using a primitive function on the
right-hand side of
Eq.\,(\ref{anti-PCF}), we find that
the phase difference obeys a gradient dynamical system
\begin{eqnarray}
\frac{d  \psi}{d t}&=& -\frac{d U}{d \psi},\\
U(\psi)&\equiv& -\int_{-\pi}^{\psi}\left[ \Delta \omega + 2K\hat{\Gamma}_a(\zeta)\right]\,  d\zeta.
\end{eqnarray}
As $t \to \infty$, the state approaches a stable point
at the bottom of the potential $U$.
The realization of synchronization is equivalent to
the existence of a phase difference $\psi_{\mathrm{sync}}$ that satisfies
\begin{eqnarray}
\frac{dU}{d\psi}&=&-\Delta \omega - 2 K\hat{\Gamma}_a(\psi_{\mathrm{sync}})=0,\label{judge1}\\
\mbox{subject to}&&\nonumber\\
\frac{d^2U}{d\psi^2}&=& -2K\hat{\Gamma}_a'(\psi_{\mathrm{sync}})>0. \label{judge2}
\end{eqnarray}
Taking the average of Eqs.\,(\ref{eq_2elem1}) and (\ref{eq_2elem2}),
we obtain the time evolution of the phase average, $\varphi=(\psi_1+\psi_2)/2$:
\begin{eqnarray}
\frac{d  \varphi}{d t}  &=& \overline{\omega} \left[ 1 + \frac{K}{\overline{\omega}}
				 \hat{\Gamma}_s(\psi) \right], \label{sym-PCF} \\
\hat{\Gamma}_s(\psi)&\equiv&\frac12\left( \hat{\Gamma}(\psi) +
				    \hat{\Gamma}(-\psi)\right), \label{def_sym}
\end{eqnarray}
where $\overline{\omega}=\omega+\left(\delta \omega_1 + \delta \omega_2 \right)/2$.
When synchronization is achieved, the frequency is shifted to
\begin{eqnarray}
\left.\frac{d \varphi}{d t}\right|_{\mathrm{sync}}&=& \overline{\omega} \left[ 1+\frac{K}{\overline{\omega}}
\hat{\Gamma}_s(\psi_{\mathrm{sync}}) \right] \label{freq_sync}.
\end{eqnarray}

We calculated the phase sensitivity,
$V^*$, with a relaxation method
\citep{ermentrout1996type,ermentrout2010mathematical}, using
a numerical integration of the
adjoint model of the Dieterich-Ruina
oscillator. 
The integration is also performed using a fourth-order Runge-Kutta
scheme with variable time-step sizes \citep{press1992numerical}.
Figure\,\ref{phase_sens} shows the phase sensitivity, or the values of $V^*$, during a time interval.
The value of the sensitivity remains positive for most of the period
except at the moment when a slip event occurs. After the slip event,
the sensitivity starts to increase for a while, following which it gradually decreases.
Using the calculated values of $V$, $\theta$, and $V^*$  as functions of phase, we have also calculated the PCF
for the oscillator according to Eq.\,(\ref{G_ij}).
Figure\,\ref{phase_coupling}(a) shows the PCF as a function of phase.
 Figure\,\ref{phase_coupling}(b) shows the PCF
on negative and  positive half-planes of phase in a logarithmic scale.
The PCF is classified as an
 anti-phase type as in Fig.\,\ref{phase_coupling_weak}, although the shape does not resemble a
 sine curve.

By checking the positional relation between the horizontal line,
$\hat{\Gamma}=-\Delta \omega/(2K)$,
and antisymmetric part, $\hat{\Gamma}_a$, of the PCF curve in Fig.\,\ref{phase_coupling},
we can determine whether Eq.\,(\ref{judge1}) subject to inequality\,(\ref{judge2}) has a solution,
i.e.,
we can evaluate whether synchronization is achieved.
In this setting, synchronization is expected in the range of 
$-6.5 \times 10^{-15} < -\Delta \omega/(2K) <
 6.5 \times 10^{-15} \mathrm{kg^{-1} m^2 s}$.
If there is an intersection between the horizontal line and
antisymmetric part, $\hat{\Gamma}_a$, of the PCF, in addition to $\hat{\Gamma}_a$ being a decreasing
function
of the phase at that point, then the
synchronization is achieved with the difference of phase at which the
intersection is located, as indicated in Fig.\,\ref{phase_coupling}(a).
The frequency of synchronized oscillators is also derived using the symmetric part,
$\hat{\Gamma}_s$, of the PCF according to Eq.\,(\ref{freq_sync}).

\subsection{Comparison of the results of numerical integration and phase reduction}
In table\,\ref{sync_prop},
important quantities representing the synchronization properties
are summarized:
the difference of the natural frequencies $\Delta \omega$,
phase difference $\psi_{\mathrm{sync}}$, and
frequency $\left.d\varphi/dt\right|_{\mathrm{sync}}$.
Data in the parentheses are
the estimated values for the synchronization properties
of the oscillator pairs, which are derived from the
intersection of the PCF and a horizontal line.
The estimated values from the PCF are in reasonable agreement with the
corresponding ones by numerical integration.
This indicates that the synchronization properties of coupled
oscillators can be quantitatively estimated using the phase reduction
method 
when
the coupling is sufficiently weak.
The PCF $\hat{\Gamma}_a$ has a pretty complicated ``micro-structure''
near $| \psi |\simeq 0$, a flat hill-like structure $0\lesssim\psi<10^{-2}$
with a sudden jump to the origin, as shown in Fig.\,\ref{phase_coupling}(b). 
Thus, we cannot decide the exact phase difference at which the oscillators are nearly
synchronized in-phase. However, we are sure that the phases never become exactly in-phase, 
where $\hat{\Gamma}'_a$ violates the inequality\,(\ref{judge2}).


According to the phase reduction method,
the range of parameters in which two oscillators synchronize is 
estimated to be
\begin{eqnarray}
\eta &\equiv&\left| 
 1.6 \times 10^{-2} \frac{\Delta\left(\frac{A}{L}\right)}{K}
-1.1 \times 10^{-2} \frac{\Delta\left(\frac{B}{L}\right)}{K}
\right|
<1, \label{sync_nond}
\end{eqnarray}
where $\Delta$ represents the difference between two oscillators.
This gives  $|\Delta B|<2.7\times 10^3 \mathrm{Nm^{-2}}$
for the parameters used here, which is consistent
 with the results of numerical integrations.

\section{Three-oscillator system}\label{sec:3-oscs}
Here, we extend the analysis to a system of three identical mutually coupled
Dieterich-Ruina oscillators.
Each oscillator in the system is assumed to be described by Eqs.\,(\ref{eq0_theta}) and
(\ref{eq1_V}) and contain the same parameter set as oscillator 1 in Section\,\ref{sec:pair}.
We consider two different coupling topologies: a periodical coupling in a ring
with spring constants $k_{12}=k_{23}=k_{31}=K$
and a non-periodical coupling in a line with $k_{12}=k_{23}=K, ~k_{31}=0$.
In terms of phase, the state of the three-oscillator
system can be characterized by the phase
differences between the oscillators,
$\psi_1 = \phi_1 - \phi_2$ and $\psi_3 = \phi_3 - \phi_2$.

\subsection{Numerical Integrations}\label{sec_direct3}
We performed numerical integrations of a discrete-time version of the
original system of differential equations for the two types of coupling patterns.
Figure\,\ref{V_3elem_gl} shows the results for the periodical coupling.
After convergence, the three oscillators share
a common phase difference of $2\pi/3$, i.e., $(\psi_1,\psi_3)\simeq(\frac23 \pi,-\frac23 \pi)$.
Figure\,\ref{V_3elem_np}
 shows the results for the non-periodical coupling.
Although the convergence is rather slow, the oscillators
gradually synchronize at phase differences near $(\psi_1,\psi_3)\simeq(7.0 \times 10^{-3},2.6)$.

\subsection{Application of the PCF}\label{sec_PCF3}
By applying the phase reduction method to the system of three identical
oscillators, the evolution of the phases can be described as
 \begin{eqnarray}
 \frac{d \phi_i}{d t} &=& \omega +
  \sum_{j \neq i} k_{ij} \hat{\Gamma} \left( \phi_i-\phi_j \right),\quad i=1,2,3, \label{eq_3elem}
 \end{eqnarray}
where $\hat{\Gamma}=\Gamma/K$.

Differences between the three equations in (\ref{eq_3elem})
give the time evolution for the phase differences as in Eq.\,(\ref{anti-PCF}).
The time evolution of the system of periodically coupled oscillators can be written as
\begin{eqnarray}
\frac{d \psi_1}{d t} &=&
\Gamma(\psi_1) -\Gamma(-\psi_1)
- \Gamma(-\psi_3) + \Gamma(\psi_1-\psi_3), \\
\frac{d \psi_3}{d t} &=&
\Gamma(\psi_3) -\Gamma(-\psi_3)
- \Gamma(-\psi_1) + \Gamma(\psi_3-\psi_1).
\end{eqnarray}
The time evolution of the system of non-periodically coupled oscillators can be written as
\begin{eqnarray}
\frac{d \psi_1}{d t} &=& \Gamma(\psi_1) -\Gamma(-\psi_1)
-\Gamma(-\psi_3),\\
\frac{d \psi_3}{d t} &=& \Gamma(\psi_3) -\Gamma(-\psi_3)
-\Gamma(-\psi_1).
\end{eqnarray}
Each three-oscillator system is thereby reduced to a two-dimensional
dynamical system for the phase differences \citep[e.g.,][]{aihara2011complex};
this two-dimensional system has a symmetry because $\psi_1$ and $\psi_3$ are interchangeable.
Similar to conditions\,(\ref{judge1}) and (\ref{judge2}) for a system of
two oscillators,
the synchronization of the three-oscillator system is expected to be realized at the
stable equilibrium points of the phase flow in the $(\psi_1,\psi_3)$-plane;
these equilibrium points emerge as intersections of the nullclines for the phase flows
(Figs.\,\ref{nullcline_per} and \ref{nonp_log}).
In the upper-right part of Fig.\,\ref{nonp_log},
the nullclines for $d\psi_1/dt=0$ and $d\psi_3/dt=0$ nearly
overlap because $\Gamma(\psi_1)$ is almost
equal to $\Gamma(\psi_3)$ owing to a rather flat region of $\Gamma$
(Fig.\,\ref{phase_coupling}(b)) they share in this range.

\subsection{Comparison of the results of numerical integration and phase reduction}
The triphase synchronization \citep[e.g.,][]{aihara2011complex} in
the periodically coupled system (Fig.\,\ref{V_3elem_gl})
is achieved because the phase oscillators exclude each other with an equal intensity
owing to the anti-phase nature of the PCF (Fig.\,\ref{phase_coupling}).
It corresponds to a stable spiral in the fourth quadrant of the phase plane 
(Fig.\,\ref{nullcline_per}).
The synchronization in the
non-periodically coupled system (Fig.\,\ref{V_3elem_np})
corresponds to one of the two stable nodes in the first quadrant of the phase plane (Fig.\,\ref{nonp_log}).
The reason for the slow convergence for the latter case is
that the orbit of
the phase differences should follow a static
 pathway along one of nearly overlapped nullclines mentioned above.
In each three-oscillator system,
the phase flow has a pair of stable equilibrium points at a symmetric
position in the phase plane with different basins of attraction.
Hence,
the convergence of the phase differences is dependent on which basin the
initial condition belongs. 

\conclusions[Conclusions]  

The Dieterich-Ruina oscillator
can be viewed as a self-sustained oscillatory system with two degrees of freedom.
This concisely describes the stick-slip motion of
a slider driven by a plate through a spring
and dashpot against a rate- and state-dependent friction.

When the bifurcation parameter $\mu = \sigma(b-a)-GV_p/(2c)-Lk$ passes through zero,
it encounters a supercritical Hopf bifurcation, and
an asymptotic analytical solution (Eqs.\,(\ref{omega0}),
(\ref{sol_SL})--(\ref{omega1})) in the weakly nonlinear regime is
available for $\mu \gtrsim 0$,
which may serve as a formula for evaluating
the recurrence intervals
of slow earthquakes
if the slip instability is sufficiently weak.

Some collective behaviors are found for a pair of weakly coupled
Dieterich-Ruina oscillators.
The two-oscillator system
that has similar weakly coupled oscillators
exhibited synchronization for some combinations of the coupling strength
and similarity of the oscillators.
Synchronization is expected
in the parameter range of inequality\,(\ref{sync_nond}).
Even though different systems of oscillators should have
different criteria, a simple model for the
earthquake generation cycle along the Nankai trough
exhibited synchronization in a similar range of $\eta < 0.35$ (Cases 1, 2, and 3 of \citet{mitsui2004simple}),
which suggests
that synchronization can occur in seismogenic process.

The synchronization is anti-phase for an identical pair; however, their
phases tend to align for
non-identical pairs with weak coupling.
The phase behavior was quantitatively estimated using the phase coupling function for
the oscillator.
It is interesting that a pair of non-identical oscillators with weak coupling can nearly cause
an in-phase  synchronization. This
suggests the possibility of sequential occurrences of 
adjacent earthquakes.

Distinct phase alignment behaviors were found for three-oscillator systems.
The system of three identical oscillators equally coupled in a ring
exhibits a triphase synchronization, 
in which they arrange themselves such that they are out-of-phase with
respect to each other by $2\pi/3$. 
In contrast, if three identical oscillators (1, 2, and 3) are equally coupled in a line
with spring constants  $k_{12}=k_{23}, ~k_{31}=0$, then oscillators 1 and 2
or oscillators 2 and 3 become nearly in-phase, while the other
remains nearly anti-phase.
The synchronization properties were quantitatively estimated using
the phase reduction method.

These results demonstrate that
synchronization should occur
between several coupled oscillators in stick-slip motion, for which
we can systematically use the phase reduction method as an analytical
tool.
In the context of seismogenic processes, 
the phase reduction method can be applied
to the analysis of observed synchronization in 
a seismogenic zone that is presumed to consist of neighboring groups of faults 
moving at similar slip rates with
mutual stress coupling \citep{scholz2010large}.
Moreover, the method is still applicable
even if the inertia is included in the model or
another friction constitutive law is adopted.
It may be of interest to examine how the phase coupling function changes
its property according to these details of the modeling.
In particular, it will be meaningful to specify the extent to which the inertia term will affect the timing of slip events.

The phase description (\ref{phi_i}) has a general form applicable
to a system with an arbitrarily large number of the oscillators described by Eqs.\,(\ref{eq0_theta}) and
(\ref{eq1_V}), as long as the oscillators are weakly coupled.
As a consequence of the anti-phase nature of the oscillator,
which is evident from the inequality\,(\ref{BF-ineq}) or from the shape of the
 PCF (Fig.\,\ref{phase_coupling_weak} or \ref{phase_coupling}), 
an irregular pattern may emerge even in a homogeneous system
with a large population of diffusively coupled oscillators. This is
where we will be able to find
the Benjamin-Feir instability developing phase turbulence \citep{kuramoto1984chemical}.

\appendix

\section{The Taylor expansion}  \label{app_taylor}  
The Taylor expansion of Eqs.\,(\ref{eq_theta}) and (\ref{eq_V}) in terms of the
deviation $\vec{u} \equiv \vec{X}-\vec{X}_0$  is as follows.

\begin{eqnarray}
\frac{d \vec{u}}{d t} &=& \mathbf{L}_0 \vec{u} + \mu \mathbf{L}_1 \vec{u}+\mathbf{M}(\vec{u},\vec{u})+\mathbf{N}(\vec{u},\vec{u},\vec{u})+\mathit{h.o.t.},\\
\mathbf{L}_0 &=&
\left[\begin{array}{cc}
-\frac{V_p}{L} & -\frac{1}{V_p}\\
\frac{V_p^3 (q^2+1)}{L^2} & \frac{V_p}{L}
\end{array}\right], \\
\mathbf{L}_1 &=&
\left[\begin{array}{cc}
0 & 0 \\
\frac{C V_p^3}{L^2} & \frac{C V_p}{L}
\end{array}\right],\\
\mathbf{M}\left( \vec{u},\vec{u} \right)
&=&
\left[\begin{array}{c}
-\frac{1}{L}u_x u_y\\
c_{xx} u_x^2 + c_{xy} u_x u_y + c_{yy} u_y^2
\end{array}\right],\\
\mathbf{N}\left( \vec{u},\vec{u},\vec{u} \right)
&=&\left[\begin{array}{c}
0\\
c_{xxx} u_x^3 + c_{xxy} u_x^2 u_y + c_{xyy} u_x u_y^2 + c_{yyy} u_y^3
\end{array}\right],\nonumber\\
&&\\
\vec{u}&=&\vec{X}-\vec{X}_0=\left[\begin{array}{c}
u_x\\
u_y
\end{array}\right],\\
c_{xx}&=&-\frac{V_p^4  (q^2+1)}{L^3},  \\
c_{xy}&=&-\frac{V_p^2  (q^2+1) (d -1)}{L^2},  \\
c_{yy}&=&-\frac{d -1}{L},  \\
c_{xxx} &=& \frac{V_p^5  (q^2+1)}{L^4},  \\
c_{xxy} &=& \frac{V_p^3  (q^2+1)(d-1)}{L^3},  \\
c_{xyy} &=& \frac{V_p  (q^2+1) (d -1)d}{L^2},\\
c_{yyy} &=&  \frac{(d-1)d}{V_p L},
\end{eqnarray}
where $\mathit{h.o.t.}$ denotes higher order terms.

\section{The trapping region}  \label{app_trap}  
The region $R$ can be constructed by bounding
it with a hexagon $H=ABCDEF$
in a $\log{\theta}$-$\log{V}$ plane, where
$A=(\log{\theta_1},\log{V_1})$,
$B=(\log{\theta_4},\log{V_1})$,
$C=(\log{\theta_4},\log{V_3})$,
$D=(\log{\theta_2},\log{V_2})$,
$E=(\log{\theta_3},\log{V_2})$,
and
$F=(\log{\theta_5},\log{V_p})$.
Using small positive values $\epsilon_i,~1\le i \le 6$, we can define the
constants for these positional coordinates as
\begin{eqnarray}
 V_1&=&V_p\left(\frac{B}{Lk}-1\right)^{-1}\left(\frac{1}{1-\epsilon_1}-1\right),\\
 V_2&=&\frac{V_p}{\epsilon_2},\\
 V_3&=&\frac{V_p}{\epsilon_2(1+\epsilon_3)},\\
 \theta_1&=&\frac{B}{V_p k}(1-\epsilon_1),\\
 \theta_2&=&\frac{L}{V_p}\left[ \frac{1}{\epsilon_2}\left(1-\frac{L
						     k}{B}
						    \right)+\frac{L
 k}{B}\right]^{-1},\\
 \theta_3&=& \frac{L}{V_p}\epsilon_2 \epsilon_5,\\
 \theta_4&=& \frac{L}{V_p}\frac{1}{\epsilon_4},\\
 \theta_5&=& \frac{L}{V_p}(1-\epsilon_6)\epsilon_5.
\end{eqnarray}
An example of the trapping region is illustrated in Fig.\,\ref{trap}.
If we assign appropriate values to $\epsilon_i$,
then all the trajectories in $R$ will be confined
within it.
To be specific,
we can set
the diagonal segments $\overline{CD}$, $\overline{EF}$, and $\overline{FA}$
to be sufficiently steep, or vertical, such that
any flows on them would be trapped.
This is derived as follows.
Slopes of the flows on a $\log{\theta}$-$\log{V}$ plane are defined as
\begin{eqnarray}
\gamma
&\equiv& \frac{\frac{d }{d t} \log{V}}{\frac{d }{d t} \log{\theta}}
= \frac{\theta \frac{d V}{d t} }{V \frac{d  \theta}{d t}}
\nonumber\\
&=&
\frac{Lk}{A+g V} \left( \frac{V_p}{V} -1 \right)
\left( \frac{1}{1 - \frac{V \theta}{L}} - 1  \right)
-\frac{B}{A+g V}.
\end{eqnarray}
Here, we assess this quantity especially
on the diagonal segments $\overline{CD}$, $\overline{EF}$, and
$\overline{FA}$.
\begin{itemize}
\item Segment $\overline{CD}$\\
     Since $\overline{CD}$ has no intersections with nullcline I, and is placed on the
     upper right side of it, the quantity
     $1/(1-V \theta/L)-1$
     has a finite negative value.
     Hence, if we assign $V$ a large value,
     $|\gamma|$ can be arbitrarily small.
     In other words, if we place the segment $\overline{CD}$ in a large $V$
     region, then the flows on it should have sufficiently gentle, or horizontal, slopes to
     be trapped in the region $R$.
\item Segment $\overline{EF}$\\
     On this segment, using $V_p \le V \le V_2$ and $ 0 <
     V\theta/L \le \epsilon_5 <1$, we find
     \begin{eqnarray}
      &&\frac{1}{A+g V_2} \le \frac{1}{A+g V} \le \frac{1}{A+g V_p},\\
      &&\frac{V_p}{V_2}-1 \le \frac{V_p}{V}-1 \le 0,\\
      &&0 < \frac{1}{1-\frac{V \theta}{L}}-1 \le \frac{1}{1-\epsilon_5}-1.
     \end{eqnarray}
      From these three inequalities, we get an estimation for the negative
slope $\gamma$:
     \begin{eqnarray}
      \gamma
      &\ge& \frac{1}{A+g V_p}\left[k L \left(\frac{V_p}{V_2}-1 \right)
				    \left( \frac{1}{1-\epsilon_5}-1
				    \right) - B \right]
     \equiv \xi_1 < -1.
     \end{eqnarray}
     The rightmost inequality is a consequence of $\mu>0$.
     If we use $\xi_1$
     as the slope of the
     segment $\overline{EF}$, then
     flows on it should have sufficiently gentle slopes to
     be trapped in the region $R$.  
\item Segment $\overline{FA}$\\
     On this segment, using $V_1 \le V \le V_p$ and $
     0<V\theta/L <1$, we find
     \begin{eqnarray}
      &&0<\frac{1}{A+g V} \le \frac{1}{A+g V_1},\\
      &&0 \le \frac{V_p}{V}-1,\\
      &&0 < \frac{1}{1-\frac{V \theta}{L}}-1.
     \end{eqnarray}
     From these three inequalities, we get an estimation: 
     \begin{eqnarray}
      0>\gamma \ge -\frac{B}{A+g V_1}
      \equiv \xi_2 <-1.
     \end{eqnarray}
     Using the same method as the $\overline{EF}$ case, if we use
     $\xi_2$
     as the slope of the
     segment $\overline{FA}$,
     then flows on it should have gentle slopes to
     be trapped in the region $R$.
\end{itemize}
See \citet{strogatz2001nonlinear}
for the construction of trapping regions.

\section{The range of application of the phase reduction method}  \label{app_app}

To investigate the range of application 
of the phase reduction method, we quantify here the weakness of heterogeneity and interaction of the oscillators
 in terms of the system of mutually coupled Dieterich-Ruina
oscillators.
Since the orbit of oscillator is well captured on a
logarithmic scale as in Fig.\,\ref{limitcycle}, 
it is convenient to deal with
the logarithm of the variables in this discussion.
The time evolution of $(\log{\tilde{\theta}_i},\log{\tilde{V}_i})$ can
be written in a dimensionless form:
\begin{eqnarray}
\frac{d\log{\tilde{\theta}_i}}{d\tau}&=&
\left( \frac{2\pi}{\omega}\frac{V_p}{L} \right)
\left(  
1-\tilde{V}_i\tilde{\theta}_i
\right), \label{logTheta_nond}\\
\frac{d\log{\tilde{V}_i}}{d\tau}&=&
\left(\frac{2\pi}{\omega}
\frac{k_i V_p}{A_i}
\right)
\frac{1}{\frac1{\tilde{V}_i}+\left(\frac{g V_p}{A_i}\right)}
\left[
1-\tilde{V}_i-\left(\frac{B_i}{k_iL_i}\right)
\left(1-\tilde{V}_i\tilde{\theta}_i\right)\frac1{\tilde{V}_i \tilde{\theta}_i}
\right]
\nonumber\\
&& +
\left( \frac{2\pi}{\omega} \frac{k_{ij} V_p}{A_i} \right)
\frac{1}{\frac1{\tilde{V}_i}+\left(\frac{g V_p}{A_i}\right)}
\left( \frac{\tilde{V}_j}{\tilde{V}_i}-1 \right),\label{logV_nond}
\end{eqnarray}
where $\tilde{\theta}=\theta V_p/L, \tilde{V_i}=V_i/V_p, \tau=\omega t/(2\pi).$

We can apply the phase reduction method
if the perturbations caused by the oscillator difference and 
the coupling term are sufficiently smaller than the absolute value of
the Floquet exponent for the amplitude mode of the limit cycle $d\vec{X}/dt=\vec{F}(\vec{X})$.
This condition ensures that the orbits of coupled oscillators 
stay in the neighborhood of the original limit cycle orbit owing to the restoring effect.
The Floquet exponent for the amplitude mode
of oscillator 1 in section \ref{sec:pair} is
 estimated to be
$\lambda=-8.8\times 10^{-10}\mathrm{s^{-1}}$,
while the averaged perturbations caused by the oscillator difference and
the coupling term are estimated to be
\begin{eqnarray}
\Delta \lambda_h &\equiv& \Delta B \frac{V_p}{AL}
\left|\frac1{2\pi}\int
\frac{-1}{\frac1{\tilde{V}}+\frac{g V_p}{A}}
\left[
\left(1-\tilde{V}\tilde{\theta}\right)\frac1{\tilde{V}\tilde{\theta}}
\right]
 d\phi \right|\nonumber \\
&=& \Delta B \times \left( 8.0\times 10^{-15}\mathrm{N^{-1}m^2s^{-1}}\right),\label{B3}\\
\Delta \lambda_c &\equiv& \frac{K V_p}{A}
\frac1{2\pi}\int \frac{1}{\frac1{\tilde{V}}+\frac{g
 V_p}{A}}d\phi \nonumber \\
&=&  K \times \left( 3.5\times 10^{-15}\mathrm{N^{-1}m^3s^{-1}}\right),\label{K2}
\end{eqnarray}
where $\Delta B=B_i-B_j$, $K=k_{ij}$, and the integrations are performed
along the limit cycle orbit.
Substituting these into  $\Delta \lambda_h, \Delta \lambda_c\ll|\lambda|$, we get the conditions for $\Delta B$ and $K$:
\begin{eqnarray}
\Delta B &\ll&  1.1\times 10^5 \mathrm{Nm^{-2}},\\
K &\ll& 2.5 \times 10^5\mathrm{Nm^{-3}}.
\end{eqnarray}

Furthermore, we can apply averaging over a period to derive the phase shifts 
if the averaged perturbation on the phase
does not alter the natural frequency substantially.
The natural frequency of oscillator 1 in section \ref{sec:pair} is 
$\omega = 1.0 \times 10^{-9}\mathrm{s^{-1}}$,
while the perturbations on the phase caused by the oscillator difference
and the coupling term are estimated, using Eqs.\,(\ref{G_ij}) and
(\ref{eq_domega}), to be
\begin{eqnarray}
\Delta \omega_h &\equiv& 
\Delta B
\left|
\frac{1}{2\pi}
\int_0^{2\pi}
\frac{-V^*(\phi)/\theta(\phi)}{A/V(\phi)+g}
\left(1-\frac{V(\phi)\theta(\phi)}{L}\right)
d\phi\right|\nonumber\\
&=&\Delta B \times \left(1.4 \times 10^{-14}\mathrm{N^{-1}m^2s^{-1}}\right), \\
\Delta \omega_c &\equiv& K\, \max_{\psi}\left|\hat{\Gamma}(\psi) \right|
\nonumber\\
&=&K\times \left(1.3 \times 10^{-14}\mathrm{N^{-1}m^3s^{-1}}\right).
\end{eqnarray}
Substituting these into $\Delta \omega_h, \Delta \omega_c\ll\omega$, we get the conditions for $\Delta B$ and $K$:
\begin{eqnarray}
\Delta B &\ll&  7.1\times 10^4 \mathrm{Nm^{-2}},\\
K &\ll& 7.6 \times 10^4\mathrm{Nm^{-3}}.
\end{eqnarray}

Taking into account these criteria,
we choose these parameters in the range of
$0 \le \Delta B \le 5\times 10^3\mathrm{Nm^{-2}}$ 
and $0 \le K \le 3\times 10^3\mathrm{Nm^{-3}}$.
We have also checked directly that each orbit in the numerical integrations stays in the
neighborhood of the original limit cycle orbit. 
Figure\,\ref{limitcycle_comp} shows
a comparison of the orbits.

\begin{acknowledgements}
We gratefully acknowledge the helpful comments and suggestions from Dr. Ralf Toenjes and an anonymous reviewer.
We are grateful to Drs. M. Toriumi and H. Noda for their helpful discussions.
We also thank Dr. Y. Hiyoshi for providing the numerical model of
 the oscillator.
\end{acknowledgements}

\bibliographystyle{copernicus}
\bibliography{./ref}
\newpage
\addtocounter{figure}{0}\renewcommand{\thefigure}{\arabic{figure}}

\begin{figure}[t]
\vspace*{2mm}
\begin{center}
\includegraphics[angle=-90,width=9.5cm]{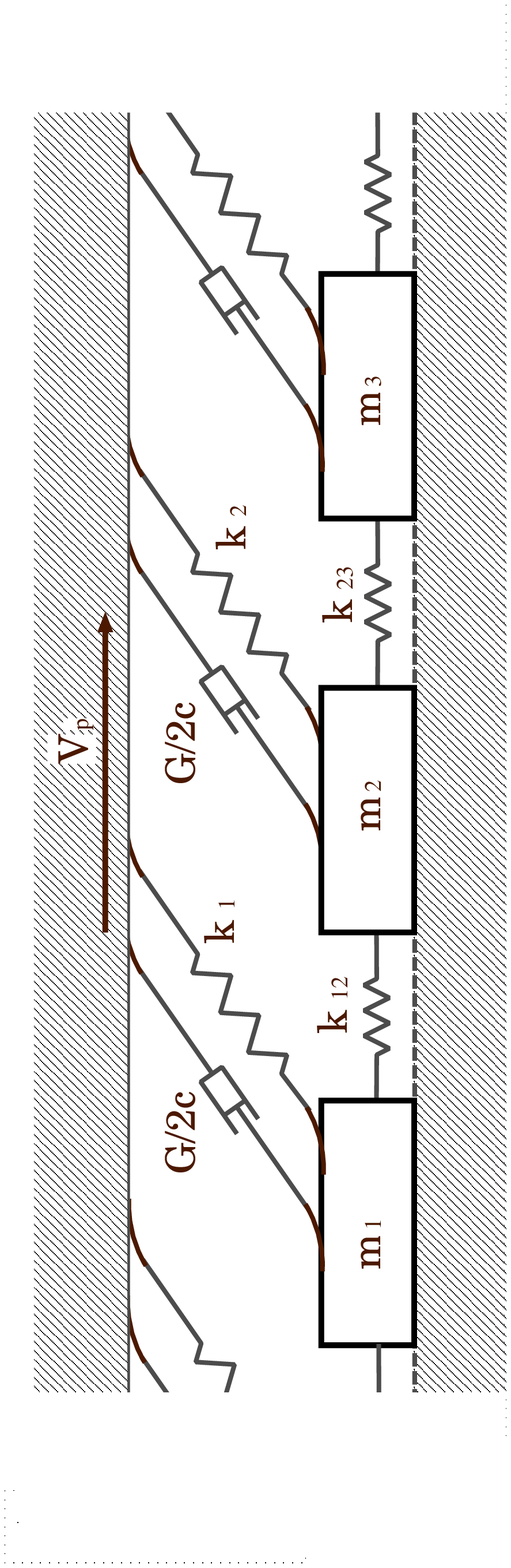}
\end{center}
\caption{Diagram of the spring-slider-dashpot system.
The configuration is identical to the Burridge-Knopoff model
 \citep{burridge1967model}, except that
it is also equipped with dashpots and the friction on the bottom of
 the sliders is rate- and state-dependent.
\label{spring}
}
\end{figure}

\begin{figure}[t]
\vspace*{2mm}
\begin{center}
\includegraphics[width=8.3cm]{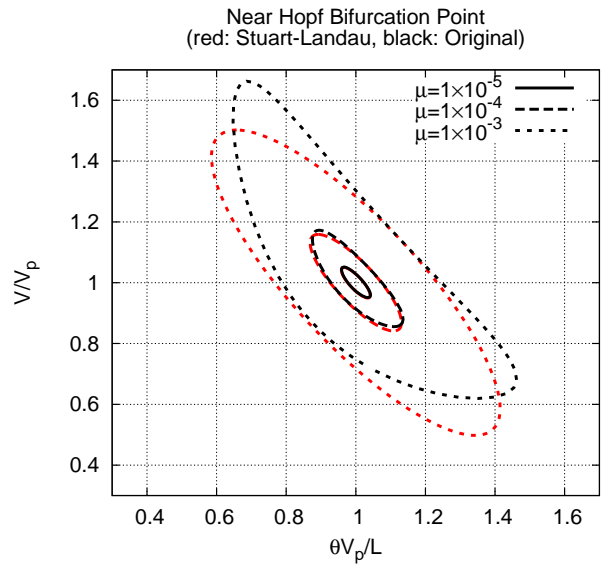}
\end{center}
\caption{Periodic orbits of the Dieterich-Ruina oscillator
near the bifurcation point, graphs of
$\left(\theta V_p/L, V/V_p \right)$.
Red and black curves are for the periodic solutions of the Stuart-Landau
 and original differential equations, respectively.
The equilibrium point is $(1, 1)$.
The values of bifurcation parameter $\mu$ are set to $1\times
 10^{-5} \mathrm{N m^{-2}} $ (solid curves),
 $1\times 10^{-4} \mathrm{N m^{-2}} $
(dashed curves), and $1\times 10^{-3} \mathrm{N m^{-2}} $ (dotted curves).
The values of $k$ used here are obtained by setting $k = (B-A-g V_p-\mu)/L$ and using the corresponding values of $\mu$.
The rest of parameters are set to
$(V_p, g, A, B, L)=
(
3.17 \times 10^{-9}\,\mathrm{ms^{-1}},
~ 5.00 \times 10^6\,\mathrm{Nm^{-3}s},
~ 1.50 \times 10^5\,\mathrm{Nm^{-2}},
~ 2.20 \times 10^5\,\mathrm{Nm^{-2}},
~ 1.00 \times 10^{-2}\,\mathrm{m})$.
\label{limitcycle0}
}
\end{figure}

\begin{figure}[t]
\vspace*{2mm}
\begin{center}
\includegraphics[width=8.3cm]{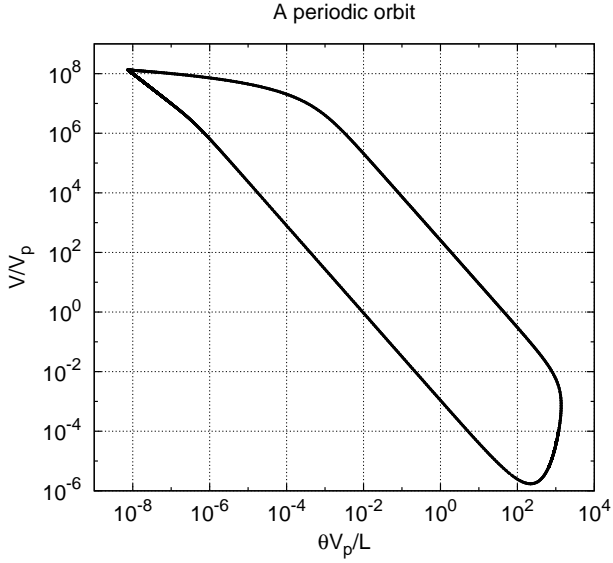}
\end{center}
\caption{A periodic orbit of the Dieterich-Ruina oscillator,    
a graph of $\left(\theta V_p/L, V/V_p \right)$,
in a double logarithmic plane.
The parameters are set to
$(k, V_p, g, A, B, L)=
(1.00 \times 10^5 \,\mathrm{Nm^{-3}},
~ 3.17 \times 10^{-9}\,\mathrm{ms^{-1}},
~ 5.00 \times 10^6\,\mathrm{Nm^{-3}s},
~ 1.50 \times 10^5\,\mathrm{Nm^{-2}},
~ 2.20 \times 10^5\,\mathrm{Nm^{-2}},
~ 1.00 \times 10^{-2}\,\mathrm{m})$.
The equilibrium point is $(1, 1)$.
\label{limitcycle}
}
\end{figure}

\begin{figure}[t]
\vspace*{2mm}
\begin{center}
\includegraphics[width=8.3cm]{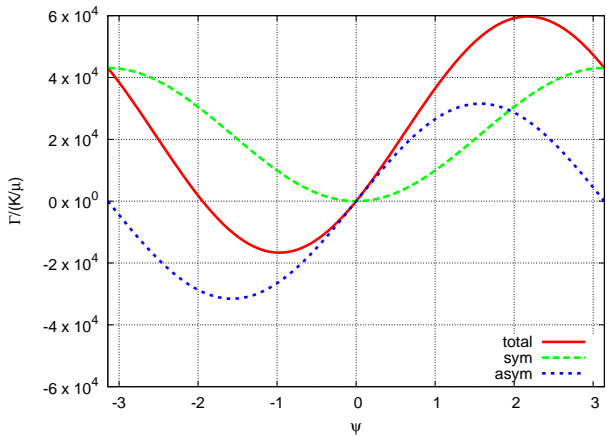}
\end{center}
\caption{The phase coupling function as a
 function of phase,
 normalized by $\mu/k_{12}=\mu/k_{21}$ for a weakly nonlinear
 oscillator. The parameters are set to
$(V_p, g, A, B, L)=
(
3.17 \times 10^{-9}\,\mathrm{ms^{-1}},
~ 5.00 \times 10^6\,\mathrm{Nm^{-3}s},
~ 1.50 \times 10^5\,\mathrm{Nm^{-2}},
~ 2.20 \times 10^5\,\mathrm{Nm^{-2}},
~ 1.00 \times 10^{-2}\,\mathrm{m})$.  
The blue, green, and red curves are the antisymmetric part defined by
 Eq.\,(\ref{def_anti}), symmetric part by Eq.\,(\ref{def_sym}),
 and total by Eq.\,(\ref{G_ij}), respectively.  
\label{phase_coupling_weak}
}
\end{figure}

\begin{figure*}[t]
\vspace*{2mm}
\begin{center}
\includegraphics[width=12cm]{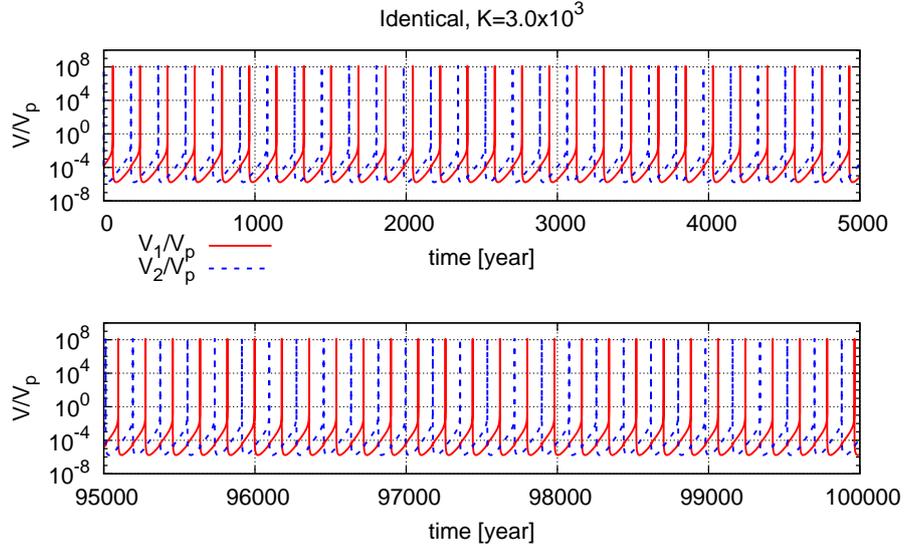}
\end{center}
\caption{The time evolution of $V/V_p$ for case 0 in a logarithmic scale.
The variation from $5000 $ to  $95000 \mathrm{yr}$ is not shown.
The parameters are
$B=2.20\times10^5 \mathrm{N m^{-2}}$ for both oscillators,
and
$K=3\times 10^3 \mathrm{N m^{-3}}$.
The oscillators synchronize at a phase difference of $\psi=-3.14$ (anti-phase).
\label{V_case0}
\addtocounter{figure}{0}\renewcommand{\thefigure}{\arabic{figure}}}
\end{figure*}

\begin{figure*}[t]
\vspace*{2mm}
\begin{center}
\includegraphics[width=12cm]{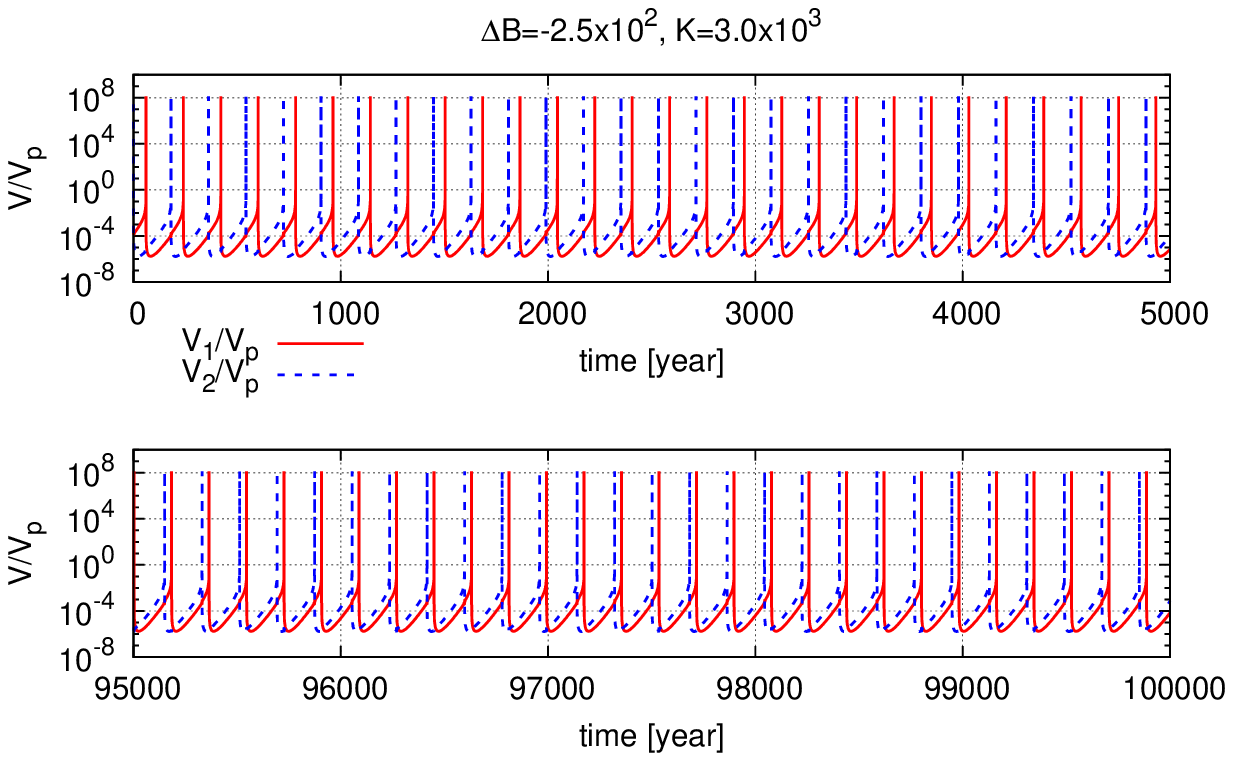}
\end{center}
\caption{The time evolution of $V/V_p$ for case 1 in a logarithmic scale.
The variation from $5000$ to  $95000 \mathrm{yr}$ is not shown.
The
 parameters are
$B=2.20\times10^5  \mathrm{N m^{-2}}$ for oscillator 1,
 $B=2.2025\times10^5  \mathrm{N m^{-2}}$ for oscillator 2,
and $K=3\times 10^3 \mathrm{N m^{-3}}$.
The oscillators synchronize at a phase difference of $\psi=-1.18$.
\label{V_case20_00006}
\addtocounter{figure}{0}\renewcommand{\thefigure}{\arabic{figure}}}
\end{figure*}

\begin{figure*}[t]
\vspace*{2mm}
\begin{center}
\includegraphics[width=12cm]{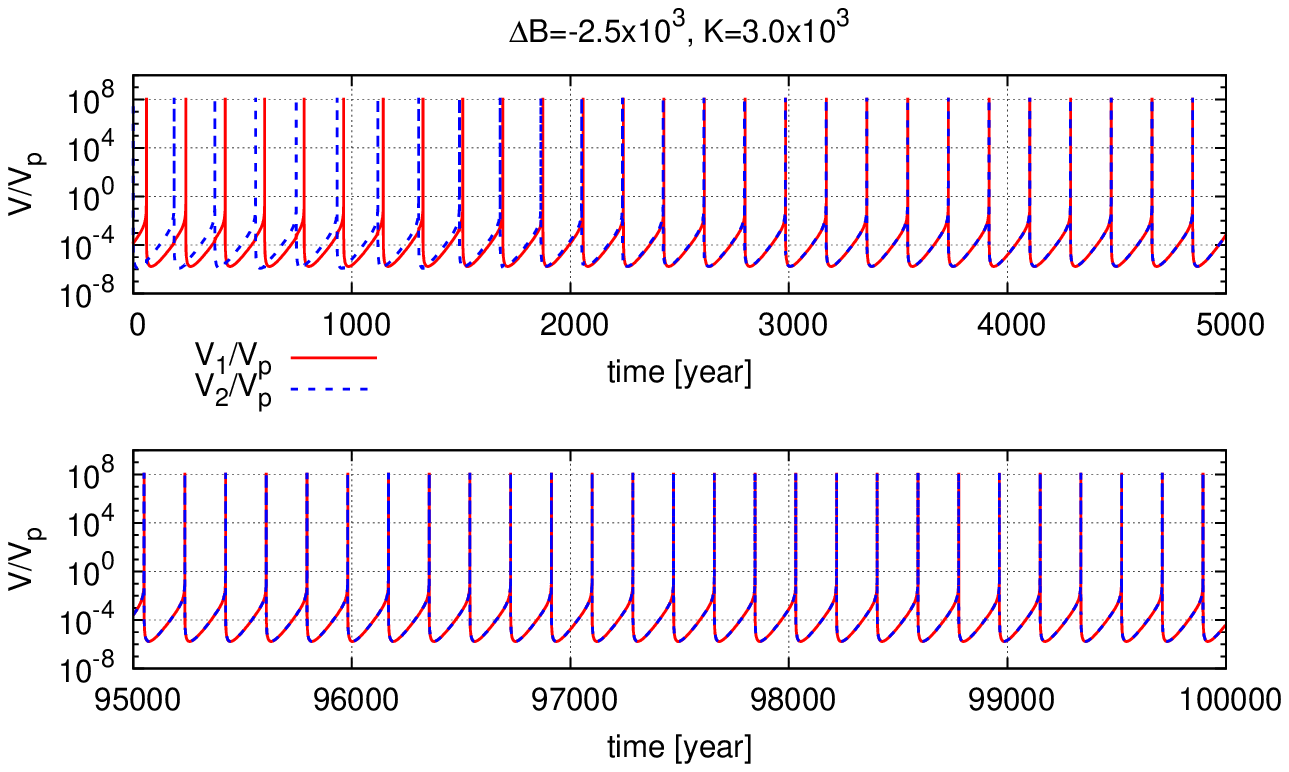}
\end{center}
\caption{The time evolution of $V/V_p$ for case 2 in a logarithmic scale.
The variation from $5000 $ to  $95000 \mathrm{yr}$ is not shown.
The
 parameters are
$B=2.20\times10^5  \mathrm{N m^{-2}}$ for oscillator 1,
 $B=2.225\times10^5  \mathrm{N m^{-2}}$ for oscillator 2,
and $K=3\times 10^3 \mathrm{N m^{-3}}$.
The oscillators synchronize at a phase difference of $\psi=-7.07 \times 10^{-3}$.
\label{V_case2_00006}
\addtocounter{figure}{0}\renewcommand{\thefigure}{\arabic{figure}}}
\end{figure*}

\begin{figure*}[t]
\vspace*{2mm}
\begin{center}
\includegraphics[width=12cm]{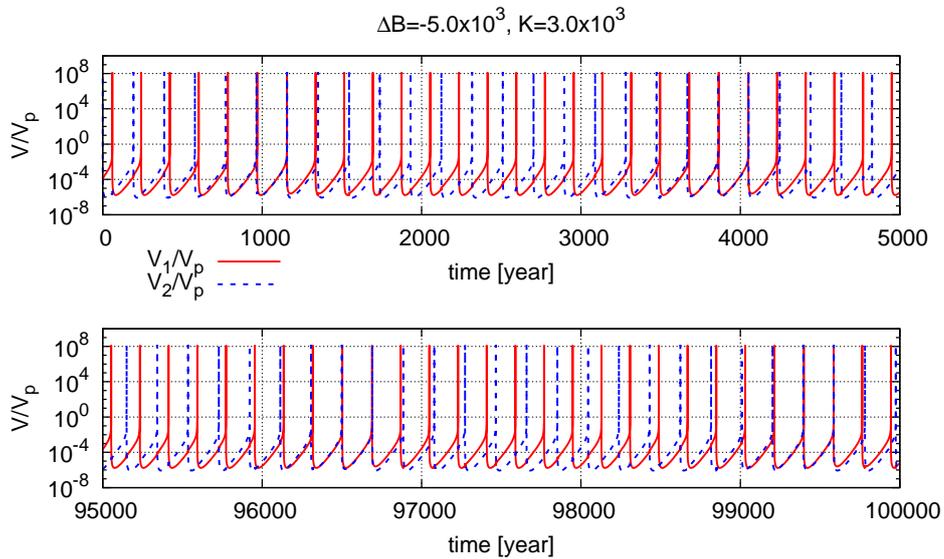}
\end{center}
\caption{The time evolution of $V/V_p$ for case 3 in a logarithmic scale.
The variation from $5000 $ to  $95000 \mathrm{yr}$ is not shown.
The
 parameters are
$B=2.20\times10^5  \mathrm{N m^{-2}}$ for oscillator 1,
 $B=2.25\times10^5  \mathrm{N m^{-2}}$ for oscillator 2,
and $K=3\times 10^3 \mathrm{N m^{-3}}$.
The oscillators are not synchronized.
\label{V_case2_00003}
\addtocounter{figure}{0}\renewcommand{\thefigure}{\arabic{figure}}}
\end{figure*}

\clearpage  

\begin{figure}[t]
\vspace*{2mm}
\begin{center}
\includegraphics[width=8.3cm]{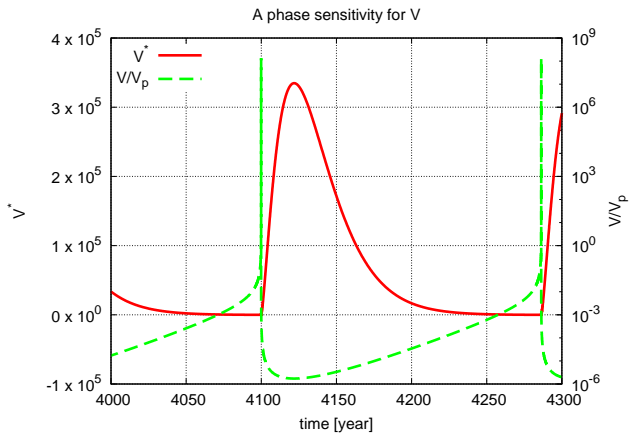}
\end{center}
\caption{Phase sensitivity, $V^*$, as a function of time in years (red
 curve)
calculated numerically using the relaxation method.
The parameters are set to
$(k, V_p, g, A, B, L)=
(1.00 \times 10^5 \,\mathrm{Nm^{-3}},
~ 3.17 \times 10^{-9}\,\mathrm{ms^{-1}},
~ 5.00 \times 10^6\,\mathrm{Nm^{-3}s},
~ 1.50 \times 10^5\,\mathrm{Nm^{-2}},
~ 2.20 \times 10^5\,\mathrm{Nm^{-2}},
~ 1.00 \times 10^{-2}\,\mathrm{m})$.
The green curve is for $V/V_p$ in a logarithmic scale.
Note that $V^*$ becomes negative for some time periods in which $V/V_p$ is large.
\label{phase_sens}
}
\end{figure}

\begin{figure}[t]
\vspace*{2mm}
\begin{center}
\hspace*{4mm}
\includegraphics[width=10.0cm]{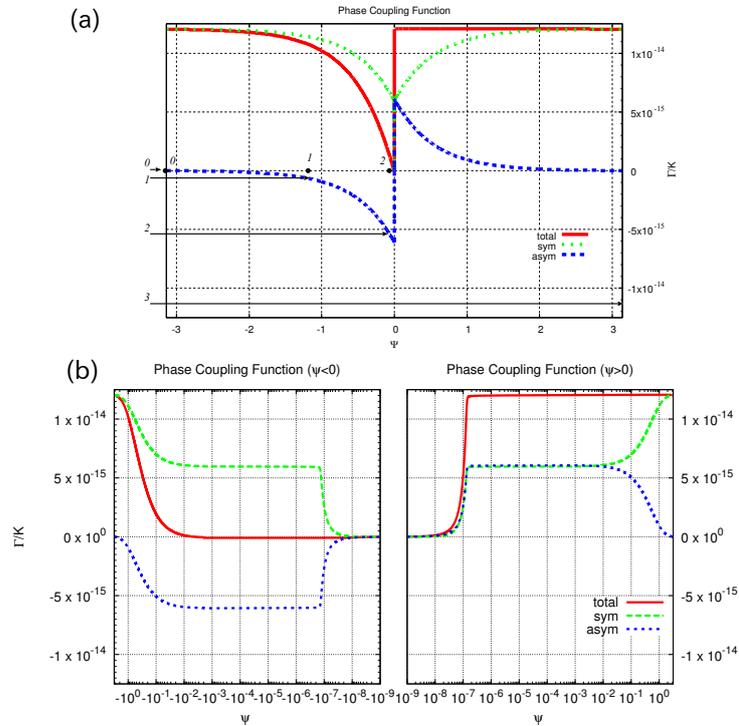}
\end{center}
\caption{The phase coupling function, $\hat{\Gamma}$, as a function of phase,
normalized by the coupling intensity $K=k_{12}=k_{21}$,
in (a) linear scale of $\psi$ and (b) logarithmic scales of $\psi$.
The parameters are set to
$(k, V_p, g, A, B, L)=
(1.00 \times 10^5 \,\mathrm{Nm^{-3}},
~ 3.17 \times 10^{-9}\,\mathrm{ms^{-1}},
~ 5.00 \times 10^6\,\mathrm{Nm^{-3}s},
~ 1.50 \times 10^5\,\mathrm{Nm^{-2}},
~ 2.20 \times 10^5\,\mathrm{Nm^{-2}},
~ 1.00 \times 10^{-2}\,\mathrm{m})$.
The blue, green, and red curves are the antisymmetric part $\hat{\Gamma}_a$,
symmetric part $\hat{\Gamma}_s$, and
total $\hat{\Gamma}$, respectively.
For each case in Table\,\ref{sync_prop},
the phase difference $\psi$ of synchronized
 oscillators and the corresponding
 value of $-\Delta \omega/(2K)$ 
  are indicated by a filled circle and an
 arrow, respectively. 
\label{phase_coupling}
}
\end{figure}

\begin{figure*}[t]
\vspace*{2mm}
\begin{center}
\includegraphics[width=12cm]{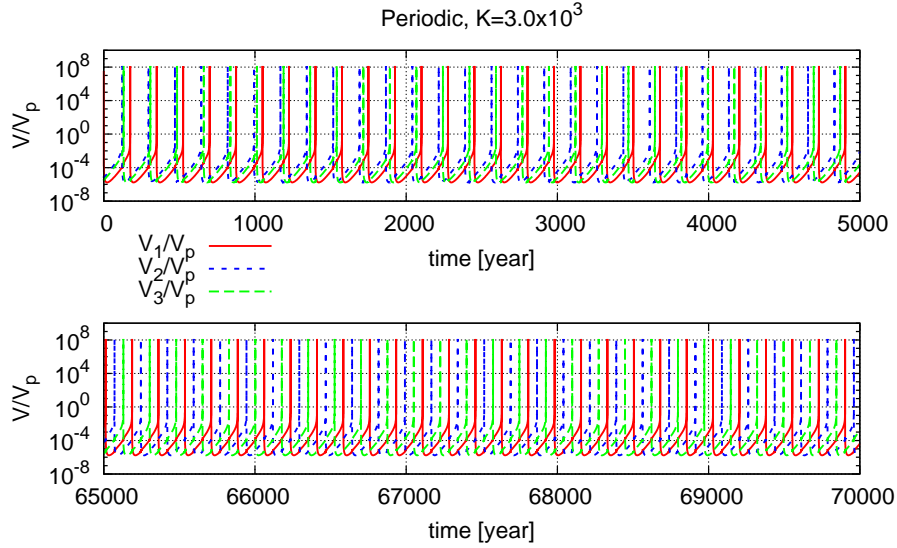}
\end{center}
\caption{The time evolution of $V/V_p$ for three identical oscillators
 with a periodic coupling in a logarithmic scale.
Red, blue, and green curves correspond to oscillator 1, 2, and 3, respectively.
The variation from $5000 $ to  $65000 \mathrm{yr}$ is not shown.
The parameters are
$B=2.20\times10^5  \mathrm{N m^{-2}}$ for all oscillators, and
$k_{12}=k_{23}=k_{31}=3\times 10^3 \mathrm{N m^{-3}}$ (periodic one-dimensional coupling).
The three oscillators synchronize at the phase differences
$(\psi_1,\psi_3)\simeq(\frac23 \pi,-\frac23 \pi)$.
This synchronization corresponds to a stable spiral in Fig.\,\ref{nullcline_per}.
\label{V_3elem_gl}
\addtocounter{figure}{0}\renewcommand{\thefigure}{\arabic{figure}}}
\end{figure*}

\begin{figure*}[t]
\vspace*{2mm}
\begin{center}
\includegraphics[width=12cm]{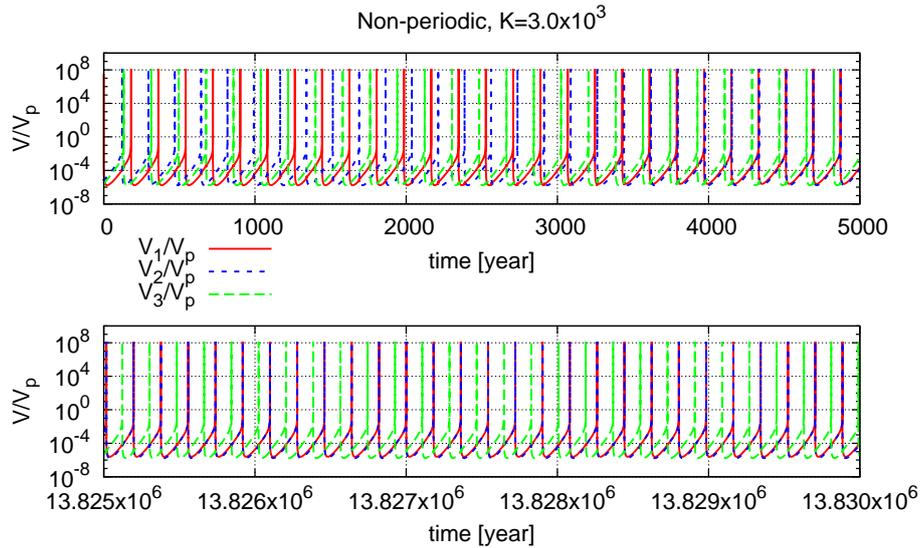}
\end{center}
\caption{The time evolution of $V/V_p$ for three identical oscillators
 with a non-periodic one-dimensional coupling in a logarithmic scale.
Red, blue, and green curves correspond to oscillator 1, 2, and 3, respectively.
The variation from $5000 $ to  $13825000 \mathrm{yr}$ is not shown.
The initial states for three oscillators are different from each other.
The parameters are
$B=2.20\times10^5  \mathrm{N m^{-2}}$ for all oscillators, and
$k_{12}=k_{23}=3\times 10^3 \mathrm{N m^{-3}}, k_{31}=0$ (non-periodic one-dimensional coupling).
They synchronize at the phase differences of
 $(\psi_1,\psi_3)\simeq(7.0 \times 10^{-3},2.6)$, where oscillators 1 and 2
are nearly in-phase.
This synchronization corresponds to a stable node in Fig.\,\ref{nonp_log}.
\label{V_3elem_np}
\addtocounter{figure}{0}\renewcommand{\thefigure}{\arabic{figure}}}
\end{figure*}

\begin{figure}[t]
\begin{center}
\includegraphics[width=7cm]{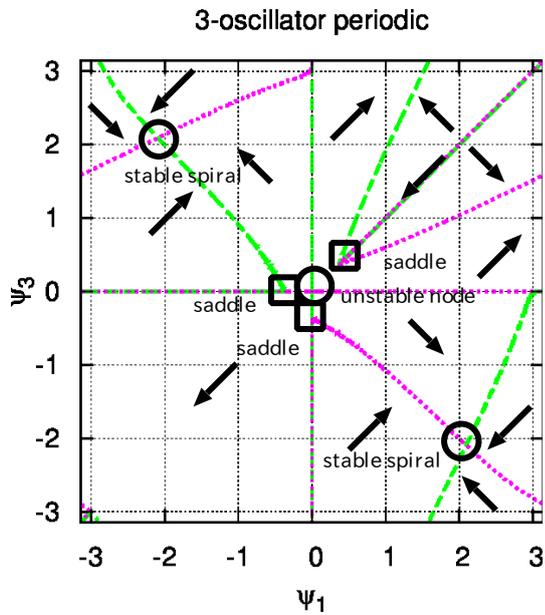}
\end{center}
\caption{Flow directions and nullclines on the $(\psi_1, \psi_3)$-plane for the phase flow of three identical
 oscillators that are periodically coupled.
Arrows indicate the flow direction.
Green and red curves represent the
 nullcline $d\psi_1/d t=0$ and
 $d\psi_3/d t=0$, respectively.
Stable spirals are located at $\left( \pm\frac23 \pi,\mp \frac23 \pi
 \right)$.
The origin is an unstable node, and the three saddles are
around $(0,-0.3)$,$(-0.3,0)$, and $(0.375,0.375)$. 
\label{nullcline_per}
}
\end{figure}

\begin{figure*}
\begin{center}
\includegraphics[width=12cm]{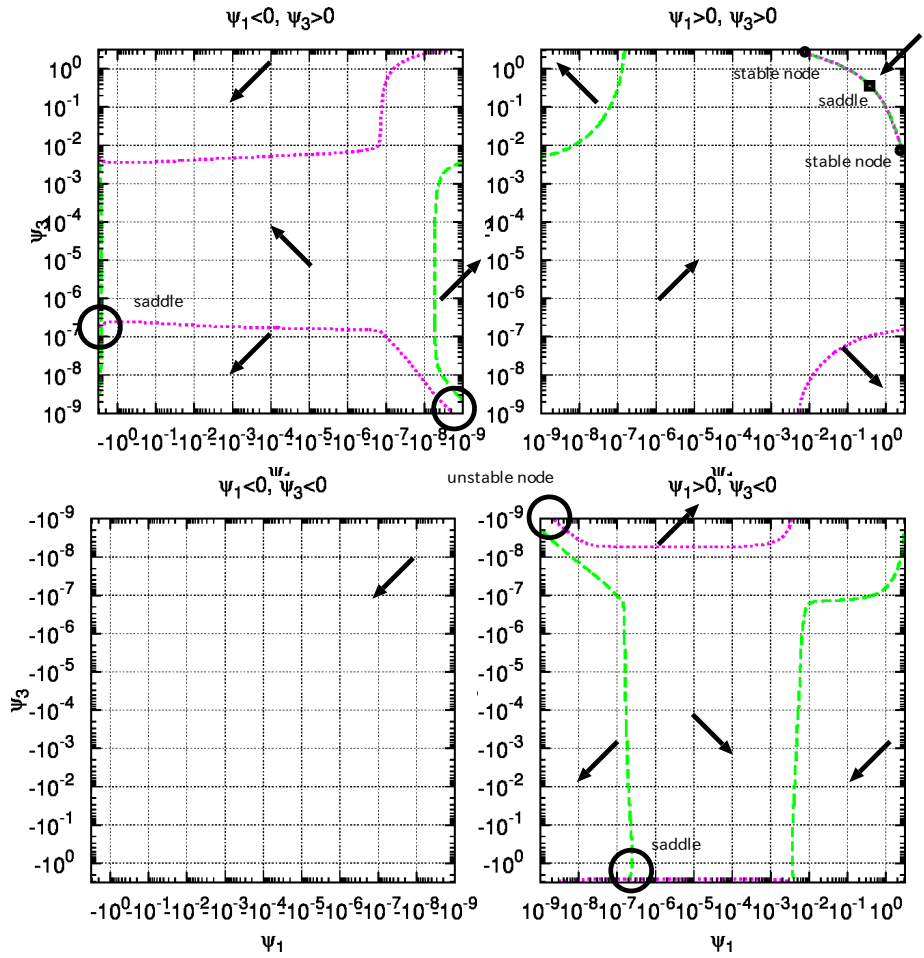}
\end{center}
\caption{Flow direction and nullclines on the $(\psi_1, \psi_3)$-plane in a logarithmic scale for the phase flow of three identical
 oscillators that are non-periodically coupled.
Arrows indicate the flow direction.
 Green and red curves represent the
nullcline  $d \psi_1/d t=0$ and
 $d\psi_3/d t=0$, respectively.
Stable nodes are located around $\left( 8 \times 10^{-3},2.6 \right)$
 and $\left( 2.6, 8 \times 10^{-3} \right)$.
The origin is an unstable node,
and saddles are around $(0.375,0.375)$,
$(2\times 10^{-7},-3)$, and $(-3,2\times 10^{-7})$.
\label{nonp_log}
\addtocounter{figure}{0}\renewcommand{\thefigure}{\arabic{figure}}
}
\end{figure*}

\begin{figure}[t]
\vspace*{2mm}
\begin{center}
\includegraphics[width=8.3cm]{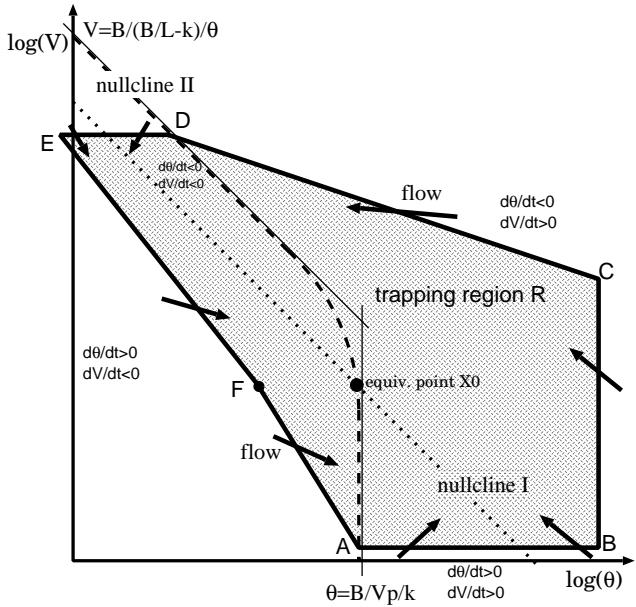}
\end{center}
\caption{A trapping region in the $\log{\theta}$-$\log{V}$ plane. Dotted line represents nullcline I, dashed curve
 represents nullcline II, black dot is the equilibrium point $\vec{X}_0$, thick solid
 line represents a hexagon $H=ABCDEF$ that bounds the
 trapping region $R$, thin solid lines represent the asymptotes for nullcline II,
 and arrows represent the flows.
\label{trap}
}
\end{figure}

\begin{figure}[t]
\vspace*{2mm}
\begin{center}
\includegraphics[width=8.3cm]{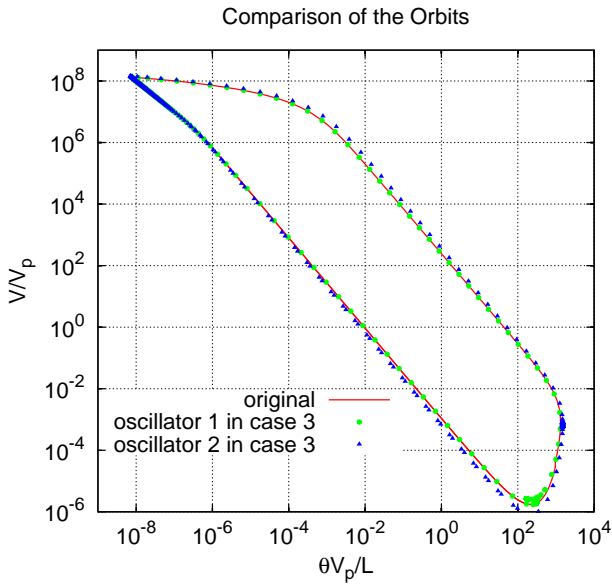}
\end{center}
\caption{A comparison of the orbits,    
graphs of $\left(\theta V_p/L, V/V_p \right)$,
in a double logarithmic plane.
The red curve, green dots, and blue triangles are the orbits of the
 original limit cycle, oscillator 1 and 2 in case 3, respectively.
\label{limitcycle_comp}
}
\end{figure}

\clearpage

\addtocounter{table}{0}\renewcommand{\thetable}{\arabic{table}}
\begin{table*}[t]
\caption{The synchronization properties of
some pairs of coupled oscillators with different parameter settings.
The corresponding values estimated from the PCF are shown in parentheses.
 \label{sync_prop}
}
\vskip4mm
\centering
\begin{tabular}{|l|r|r|r|r|}
\tophline
 Case & 0& 1& 2& 3\\
\hline \hline
$K$ ~~$[\mathrm{Nm^{-3}}]$ &
\multicolumn{4}{r|}{ $3.0 \times 10^3$ }\\
\hline
\multirow{2}{*}{$\Delta \omega$ ~~ $[\mathrm{s^{-1}}]$ }  &
0&
$3.57 \times 10^{-12}$&
$3.47 \times 10^{-11}$&
$6.73 \times 10^{-11}$
\\
&(0)&
($3.59 \times 10^{-12}$)&
($3.59 \times 10^{-11}$)&
($7.19 \times 10^{-11}$)
\\
\hline
\multirow{2}{*}{$-\frac{\Delta \omega}{2K}$ ~~ $[\mathrm{kg^{-1}m^2s}]$ }  &
0&
$-5.96 \times 10^{-16}$&
$-5.79 \times 10^{-15}$&
$-1.12 \times 10^{-14}$
\\
&(0)&
($-5.99 \times 10^{-16}$)&
($-5.99 \times 10^{-15}$)&
($-1.19 \times 10^{-14}$)
\\
\hline
\multirow{2}{*}{Synchronized?}
&  Yes
&  Yes
&  Yes
&  No
\\
& (Yes)
& (Yes)
& (Yes)
& (No)
\\
\hline
\multirow{2}{*}{$\psi_{\mathrm{sync}}$}
& $-3.14$
& $-1.18$
& $-7.07 \times 10^{-3} $
& -
\\
& ($-3.14$)
& ($-1.23 $)
& ($-7.53 \times 10^{-3}$)
&  ( - )
\\
\hline
\multirow{2}{*}{$\left.\frac{d \varphi}{d
 t}\right|_{\mathrm{sync}}-\overline{\omega}$ ~~ $[\mathrm{s^{-1}}]$ }
& $3.62 \times 10^{-11} $
& $3.44 \times 10^{-11} $
& $1.76 \times 10^{-11} $
& -
\\
& ($3.61 \times 10^{-11} $)
& ($3.44 \times 10^{-11} $)
& ($1.81 \times 10^{-11} $)
& ( -  )
\\
\bottomhline
\end{tabular}
\end{table*}

\end{document}